\documentclass[12pt,a4paper]{article}
\usepackage{amsfonts}
\usepackage{amsmath}
\usepackage{bm}
\usepackage{fancyhdr}
\usepackage[a4paper, total={6.5in, 10in}]{geometry}
\usepackage{amssymb}
\usepackage{float}
\usepackage{caption}
\usepackage{subcaption}
\usepackage{graphicx}
\usepackage{natbib}
\usepackage{hyperref}
\usepackage{booktabs}
\usepackage{chngcntr}
\usepackage{placeins}
\usepackage{authblk}%
\usepackage{rotating}
\usepackage{tikz}
\usetikzlibrary{positioning}
\usepackage{enumitem}
\usepackage{grffile}

\providecommand{\U}[1]{\protect\rule{.1in}{.1in}}

\newtheorem{theorem}{Theorem}

\newtheorem{proposition}[theorem]{Proposition}

\newenvironment{proof}[1][Proof]{\noindent\textbf{#1.} }{\ \rule{0.5em}{0.5em}}

\hypersetup{
  colorlinks   = true, 
  urlcolor     = blue, 
  linkcolor    = blue, 
  anchorcolor = blue,
  citecolor   = blue 
}

\begin{document}

\title{The addicted predator–prey model: How opioid use disorder shapes productivity and growth-cycle dynamics\thanks{An earlier version of this paper was be presented at the Keynes Year Conference, Mexico City, Mexico; and will be presented at the ASSA 2027 Annual Meeting, Washington DC, United States. We are grateful to the participants for their insightful comments and suggestions. Special thanks are due to Brad Conner, Daniele Tavani, and Ray Miller for the pleasant discussions during the preparation of an earlier draft of the article. The content is solely the responsibility of the authors, and usual caveats apply.}}

\author{Nara Chung$^{\dagger}$ \: $\cdot$ \: Marwil J. Dávila-Fernández$^{\dagger}$}

\date{{\normalsize {$^{\dagger}$\emph{Colorado State University}}} \bigskip \bigskip \\
August 2026}


\maketitle

\begin{abstract}
This paper extends Goodwin's (1967) predator-prey growth-cycle model to incorporate the negative impact of Opioid Use Disorder (OUD) on labor productivity. Using U.S. state-level data for 1998--2019, we first document that deteriorations in the labor share are followed by rising drug-induced mortality. We then build this link into the model through a novel discrete-choice mechanism in which the probability of OUD decreases with the wage share, so when real wages fail to keep pace with productivity growth, opioid use rises. This behavioral health channel feeds back into the economy through a productivity damage function. Applying the existence part of the Andronov-Hopf bifurcation theorem, we show that the resulting three-dimensional nonlinear system admits a limit cycle, as confirmed by our numerical simulations. Along the cycle, rising OUD paradoxically raises the employment rate, as falling productivity requires more labor per unit of output; across steady states, however, higher OUD prevalence is associated with a lower employment rate, consistent with evidence that opioid use disorder reduces labor force participation. Moreover, our numerical experiments show that OUD sensitivity to income widens fluctuations, while productivity sensitivity to OUD counterintuitively compresses them through a self-correcting feedback that speeds the recovery of the wage share. Policies that weaken the link between income and OUD, such as stronger social safety nets, stabilize the resulting dynamics.

\bigskip

 \bigskip\ 


\end{abstract}

\newpage

\section{Introduction}

The sustained increase in drug-induced mortality across the United States since the late 1990s has escalated into a crisis that can no longer be treated as solely a matter of individual behavior or public-health policy. By the late 2010s, drug overdose deaths exceeded 70,000 annually, with opioids involved in the majority of fatalities. The economic cost of the opioid crisis has been estimated at more than 500 billion USD per year, or about 2.8 percent of GDP, combining healthcare expenditures, criminal justice costs, and productivity losses \citep{CEA2017}. The epidemic is reshaping the composition and productivity of the U.S. workforce in ways likely to carry persistent socio-economic consequences.

While the damaging effects of Opioid Use Disorder (OUD) on labor productivity have been documented in the literature, less appreciated is that they are endogenous to the operation of the economic system and shape macroeconomic dynamics over time. Standard macro models typically decompose such effects into separable trend and cyclical components. We argue that this decomposition may be misleading because OUD affects both dimensions at once. It erodes the workforce's productivity trend \citep{Florence2016, CheungMarchandMark2023, GhimireFlorenceGuy2026}; at the same time, it is bound up with the cycle, as labor-market slack raises the incidence of OUD, and the latter slows down workers' re-entry into employment \citep{CarpenterMcClellanRees2017, AzagbaEtAl2021, Musse2025, MukherjeeEtAl2025}. Such elements motivate our adoption of the \citet{goodwin1967} predator-prey growth-cycle model as a starting point. It allows growth and fluctuations to emerge together, placing income distribution at center stage, which matters for our purposes because distribution mediates entry into and exit from employment, wage bargaining, and productivity growth alike.

Goodwin's model remains a powerful system for doing macrodynamics, with recent contributions incorporating real-financial market interactions and the possibility of financial crisis \citep{AsadaEtAl2011, SordiVercelli2014, GiraudGrasselli2021}, political cycles \citep{DiGuilmiEtAl2023}, uneven development in open economies \citep{LaMarca2010, DavilaFernandez2020}, innovation and secular stagnation \citep{TavaniZamparelli2015, TavaniZamparelli2020}, and the economics of
climate change \citep{ScaliaEtAl2020, SordiDavilaFernandez2023, YanovskiEtAl2024}, among other extensions \citep{BarralesEtAl2022, BasuMartins2026}. We innovate by introducing OUD into this framework through a discrete-choice mechanism \citep{BrockHommes97, BrockDurlauf2001, FrankeWesterhoff2017} in which agents switch between opioid use and non-use in response to the real wage-to-labor-productivity ratio, i.e., the wage share. Motivated by the evidence presented in this paper, we assume that the probability of OUD is a decreasing function of the wage share, since a real wage low relative to productivity reflects greater worker vulnerability. This behavioral choice feeds back into the economy through a productivity damage function, which we refer to as the income-protective channel.

The model's central assumption that a reduction in the wage share raises the incidence of OUD has, to our knowledge, not previously been examined empirically. We therefore begin by establishing it directly in U.S. state-level data for 1998--2019. We use age-adjusted drug-induced mortality as a proxy for OUD. We classify annual labor-share changes into nine severity bins, from large contractions through no-change years to large expansions, and trace their relationship to our variable of interest using event-style local projections \citep{Teulings2013}. Our findings indicate that states undergoing severe labor-share contractions record drug mortality well above the sample mean, and states with labor-share gains record mortality below it. The dynamic projections sharpen this into a directional relationship over time. Labor-share gains are followed by reductions in drug-induced mortality at horizons of one to four years. We read this as first-pass evidence for the income-protective channel the model formalizes.

In our model, the same relationship takes different forms along the cycle and across steady states. Distinguishing them is important. \textit{Along the cycle}, a low wage share raises the employment rate. This happens because profit rates improve, increasing capital accumulation; OUD also rises, lowering labor productivity growth through our damage function, so sustaining a given output level requires proportionally more labor. The cost of OUD registers in productivity growth, not in the employment rate, which rises. \textit{Across steady states}, by contrast, a higher equilibrium OUD is associated with a lower employment rate, consistent with evidence on labor-force participation that we review in Section 2.2. The model thus reconciles a paradoxical cyclical comovement with the standard equilibrium relationship, precisely the kind of joint information that separating trend from cycle would discard.

Applying the existence part of the Andronov-Hopf bifurcation theorem, we show that the resulting three-dimensional system generates endogenous cycles. As a key parameter crosses its critical value, the equilibrium loses stability, and a branch of periodic solutions emerges. Our simulations confirm that these fluctuations persist and remain bounded well away from the bifurcation point. Within this regime, the system displays a weak form of path dependence. For the same parameter configuration, multiple orbits coexist, so that the cycle the economy settles into is determined by its initial position. The path dependence is weak in the sense that the coexistence is confined to periodic attractors.\footnote{For instance, the system exhibits neither chaotic attractors, which would introduce trajectories that are sensitive to initial conditions, nor coexistence across attractor types, such as a stable equilibrium alongside a limit cycle, or periodic alongside aperiodic attractors. Multiplicity is limited to periodic attractors, and the role of initial conditions is correspondingly restricted to selecting among them.}

In terms of parameter configurations, two results are worth stressing. The first concerns the sensitivity of OUD behavior to income fluctuations. Raising it widens the amplitude of the limit cycles, while lowering it has the opposite effect. We interpret lower values of this parameter as reflecting stronger social safety nets, universal healthcare access, or expanded mental-health services that stabilize the system. The second result concerns the damage function's sensitivity to OUD. One would expect a more damaging productivity response to destabilize the system. Instead, we show that a higher value strengthens a self-correcting feedback loop, in which OUD-induced productivity losses trigger a faster recovery of the wage share, ultimately compressing the cycle and eliminating the coexistence of multiple attractors. The two levers are therefore opposed. Behavioral sensitivity amplifies the fluctuations, whereas damage sensitivity both compresses them and removes their dependence on initial conditions.

The remainder of the paper is organized as follows. Section~2 reviews the empirical literature on drug abuse, labor productivity, and labor force participation, before presenting our empirical exercise. Section~3 develops the theoretical model, demonstrates the existence of an internal equilibrium solution, characterizes its local stability, and applies the existence part of the Andronov-Hopf bifurcation theorem. In Section~4, we perform a series of numerical experiments to illustrate the model's main properties. Section~5 offers concluding remarks.

\section{An overview of the literature}

This section situates our contribution within two strands of empirical research before presenting our own evidence. We review the literature linking substance abuse, including opioid use disorder, to labor productivity, turning then to its effects on labor force participation. Building on both, we present our empirical exercise, which documents the relationship between the labor share and drug-induced mortality and motivates our theoretical model.

\subsection{Substance use and labor productivity}

A broad empirical literature establishes that drug abuse reduces labor productivity through impaired physical and cognitive functioning, absenteeism, presenteeism, long-term disability, and premature mortality. These channels lower effective labor input in two distinct ways. First, by reducing output per hour worked among those who remain employed. Second, by permanently removing embodied human capital from the economy through death or exit. \citet{Florence2016} provide early U.S. evidence on the combined scale of these channels, estimating that prescription opioid overdose, abuse, and dependence imposed a total economic burden of \$78.5 billion in 2013. Of this total, roughly \$20 billion is attributable to nonfatal lost productivity, which combines reduced working hours, disability, and lost output among incarcerated individuals. A further \$21 billion reflects the lost lifetime productivity of those who died. Productivity losses include both market labor and household production, suggesting the economic damage extends beyond formal employment relationships.

More recent estimates by \citet{GhimireFlorenceGuy2026} update this picture for morbidity specifically. Substance-use-disorder-related productivity losses among U.S. adults reached \$92.65 billion in 2023, with inability to work accounting for nearly half of the total. Morbidity-related losses capture the annual value of production forgone because a disorder reduces an individual's capacity to work while alive through absenteeism, presenteeism, and inability to work. The authors exclude premature mortality, and this distinction matters because morbidity losses are recurring and, in principle, reversible. This makes morbidity a natural channel through which health deterioration affects growth dynamics, operating as an endogenous productivity wedge rather than a one-time level shock. 

Mortality-based estimates, by contrast, represent an upper bound on productivity losses, since premature death permanently eliminates all future labor contributions rather than temporarily reducing capacity among the living. Using Canadian data, \citet{CheungMarchandMark2023} estimate that opioid overdose deaths between 2016 and 2019 generated \$8.8--\$10.9 billion in lost future labor productivity, or roughly \$0.6-\$0.7 million per death under a human capital approach. More than two-thirds of overdose victims had been employed within five years of their death. Such evidence contradicts the view that opioid-related harm is concentrated among workers already at the margins of the labor market.

The occupations most exposed to opioid mortality also show how impairment operates before death occurs. \citet{Lundstrom2025} document that stimulant co-involvement in synthetic opioid overdose deaths is especially prevalent in physically demanding occupations such as construction, extraction, and agriculture, consistent with workers using stimulants to sustain alertness and output despite opioid-induced impairment. The same occupational data also show that overdose deaths are concentrated in labor-intensive sectors, where output depends most directly on the physical capacity of the workforce. Overall, the literature provides empirical support for modeling substance abuse as an endogenous drag on labor productivity.

\subsection{Drug abuse and labor force participation rate}
A closely related literature turns from productivity to participation, documenting that drug abuse reduces labor force participation, particularly among prime-age workers, by increasing disability, worsening health, and weakening labor market attachment. \citet{Krueger2017} was among the first to establish this link directly. Nearly half of prime-age men not in the U.S.\ labor force take pain medication on a given day, most commonly prescription opioids, and labor force participation declined more sharply in counties with higher opioid prescribing rates. This is consistent with health-driven exit from the labor force rather than prolonged spells of unemployment.

Quantifying these effects during the COVID-19 period, \citet{GreenwoodGunerKopecky2022} estimate that increased substance abuse accounts for 9--26 percent of the decline in prime-age labor force participation between February 2020 and June 2021. Individuals with opioid or methamphetamine use disorders have participation rates 13--16 percentage points below those of non-users. The same mortality and occupational evidence introduced in Section~2.1 speaks to participation as well as productivity: \citet{CheungMarchandMark2023} show that opioid overdose deaths in Canada disproportionately affected working-age individuals who were strongly attached to the labor market prior to death, implying permanent, non-reversible participation losses rather than transitory exits; and \citet{Lundstrom2025}'s occupational patterns indicate that overdose deaths are concentrated in physically demanding, labor-intensive occupations, suggesting that participation losses from OUD are not evenly distributed across the workforce but concentrated in specific sectors.

Beyond aggregate labor-market conditions, the composition of local growth also matters for participation. Using a Bartik-style identification strategy, \citet{BetzJones2018} show that employment growth in lower-wage industries, which disproportionately employ less-educated workers, is associated with reductions in opioid overdose mortality, while employment growth in higher-wage industries is associated with increases in overdose deaths. Their findings suggest that economic growth is not uniformly protective and that shifts in income distribution and job structure can alter substance use risk across populations. As such, labor market dynamics influence substance abuse through relative income and opportunity channels, rather than through aggregate conditions alone. Causal evidence from \citet{Beheshti2023} reinforces the same channel from the opposite direction. Exogenous reductions in opioid availability increased labor force participation through employment gains. We read the existing literature as supporting treating labor force participation as an endogenous outcome shaped by OUD.



\subsection{Linking the labor share to drug-induced mortality}

The model presented in the next Section has as a critical assumption that a reduction in the wage share raises the incidence of OUD. To our knowledge, this specific claim has not been previously examined empirically. Thus, we offer insights in that direction by assessing how labor share shocks are distributed across U.S.\ states and time, and how different levels of shock severity relate to drug-induced mortality. The analysis draws on a state-level panel covering all 50 U.S.\ states over 1998--2019. The outcome variable is the age-adjusted drug-induced mortality rate per 100,000 population. Labor share shocks are constructed as year-over-year changes in the state-level labor share, standardized by their cross-sectional standard deviation, and classified into nine bins: a no-shock category (Bin 0) for changes within $\pm 0.5$ standard deviations, four positive shock bins (Bins 1--4) for increases of $+0.5$ to $+1.0$, $+1.0$ to $+1.5$, $+1.5$ to $+2.0$, and above $+2.0$ standard deviations, and four negative shock bins (Bins 5--8) for symmetric decreases. This classification captures both the direction and severity of labor share movements and enables a dose-response analysis across shock intensities.

\begin{figure}[tbp]
    \centering
    \includegraphics[width=0.95\textwidth]{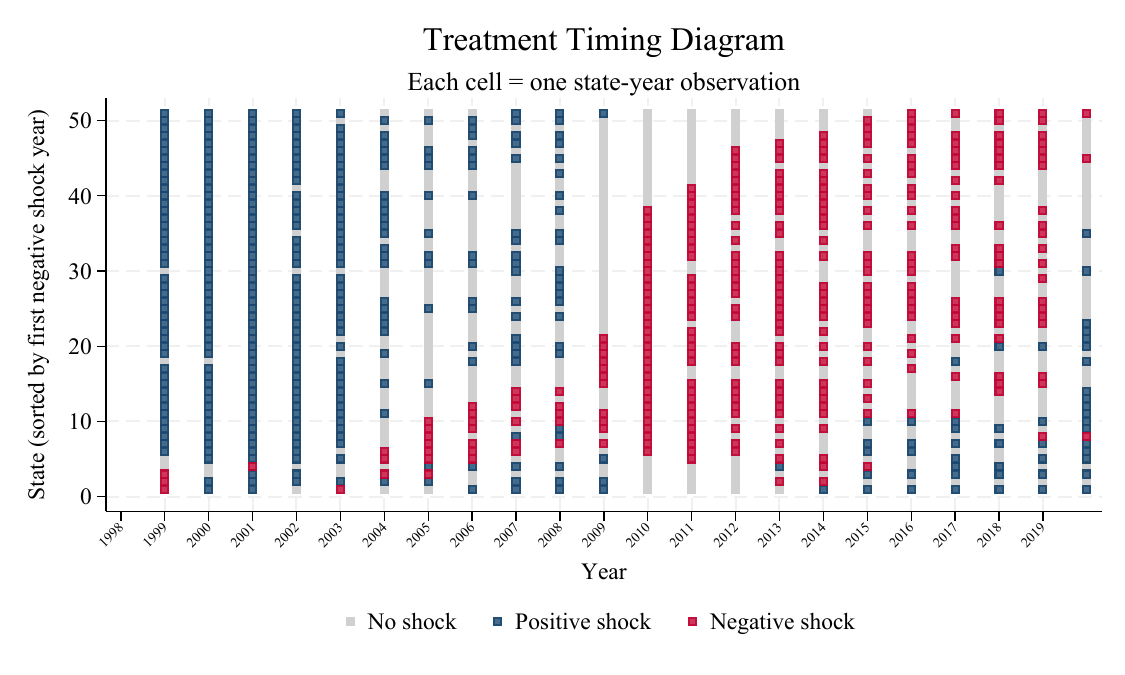}
    \caption{Treatment timing of labor share shocks across U.S.\ states,
    1998-2019. Each row is a state, sorted by year of first negative shock. Gray
    cells indicate no-shock years (within $\pm 0.5$ SD); navy cells are positive
    labor share shocks (Bins 1-4); red cells are negative labor share shocks
    (Bins 5-8). The concentration of red cells after 2008 reflects the widespread
    labor share compression associated with the Great Recession.}
    \label{fig:timing}
\end{figure}

\paragraph{Timing of labor share shocks.}
Fig.~\ref{fig:timing} maps the timing of labor share shocks for each U.S.\ state across the full sample period. Each row corresponds to a state, sorted by the year of its first negative shock. Each cell is colored gray for no-shock years, navy for positive shocks, and red for negative shocks. Two broad patterns emerge. First, positive shocks are more prevalent in the late 1990s and early 2000s, coinciding with the era's expansion and tight labor markets. Second, negative shocks become substantially more widespread after 2007-2008, reflecting the labor share compression associated with the Great Recession. By 2009, most states had experienced at least one negative labor share shock. A third feature is the recurring nature of treatment. Most states appear in multiple shock categories across different years, alternating between positive and negative episodes within the sample window. This pattern of staggered and recurring shocks motivates the use of Teulings-Zubanov local projections rather than a standard single-event-study
design.

\begin{figure}[tbp]
    \centering
    \includegraphics[width=0.85\textwidth]{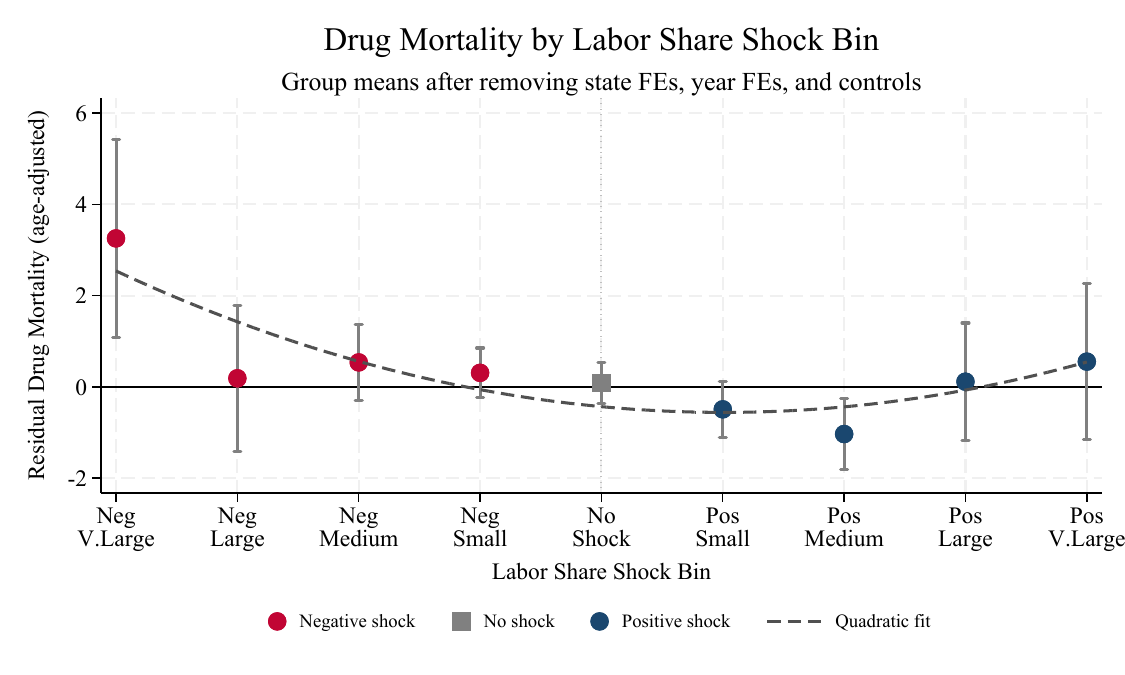}
    \caption{Mean residual log drug-induced age-adjusted mortality by labor share
    shock bin, after removing state fixed effects, year fixed effects, manufacturing
    employment share, and the unemployment rate. Bins 5-8 correspond to negative
    shocks of increasing severity; Bin 0 is the no-shock baseline; Bins 1-4 are
    positive shocks of increasing severity. The fitted curve is a quadratic
    polynomial.}
    \label{fig:groupmeans}
\end{figure}

\paragraph{Static relationship between shock severity and drug mortality.}
Fig.~\ref{fig:groupmeans} plots mean residual log drug-induced mortality by shock bin, after removing state fixed effects, year fixed effects, and cyclical controls (manufacturing employment share and the unemployment rate). These residuals capture the within-state, within-year variation in drug mortality associated with each shock category, net of aggregate trends and observable confounders. The diagram reveals a broadly monotone dose-response relationship. States in the most severe negative shock category (Bin 8, labor share declines exceeding 2 standard deviations) exhibit residual drug mortality approximately 3.3 log points above the sample mean, while states experiencing moderate positive shocks (Bin 2, increases of 1.0-1.5 SD) show mortality roughly 1.0 log point below average. Intermediate negative shock bins (Bins 5-7) are associated with modestly positive residuals, while positive shock bins (Bins 1-3) lie below zero. A fitted quadratic curve captures the overall shape. Drug mortality declines as shocks move from the most severe negative to the positive range, though the Bin 8 estimate carries little weight given only three treated state-year observations. This cross-bin pattern is consistent with the model's central prediction that labor income compression raises the incidence of substance abuse, while labor income gains are protective.

\paragraph{Dynamic effects: Local projection impulse responses.}
To assess the dynamic causal effects of labor share shocks on drug mortality, we estimate event-style local projections following \citet{Teulings2013} for each shock bin separately. Each regression controls for state and year fixed effects, the full set of nuisance shock-bin indicators (all bins other than the focal bin), manufacturing employment share, and the unemployment rate. The outcome is the cumulative log change in age-adjusted drug mortality relative to the year before the shock ($t = -1$). Pre-period estimates ($t = -6$ to $t = -2$) test for differential pre-trends, while post-period estimates ($t = 0$ to $t = +5$) recover the dynamic causal effect.

\begin{figure}[tbp]
    \centering
    \includegraphics[width=1\textwidth]{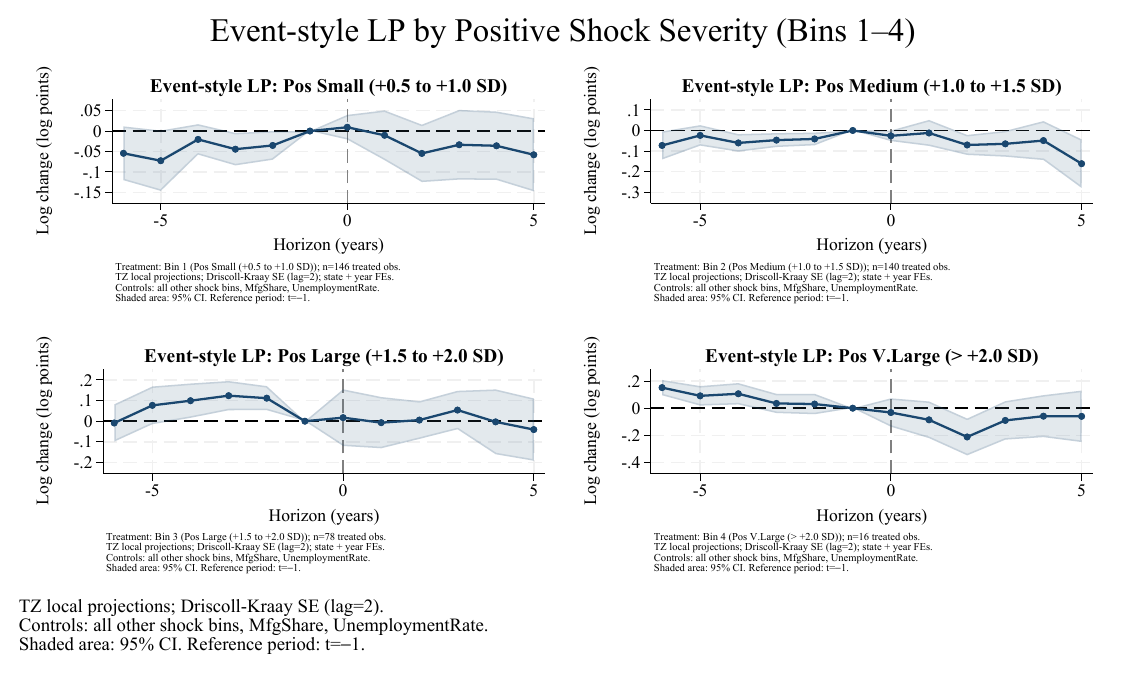}
    \caption{Event-style local projection impulse responses for positive labor share
    shock bins (Bins 1--4). The outcome is the cumulative log change in age-adjusted
    drug mortality relative to $t = -1$. Shaded areas are 95\% confidence intervals
    based on Driscoll--Kraay standard errors (lag 2). All regressions include state
    and year fixed effects, manufacturing employment share, unemployment rate, and
    all other shock bins as nuisance controls. The vertical dashed line marks the
    shock year; the horizontal dashed line marks zero.}
    \label{fig:lp_pos}
\end{figure}

Fig.~\ref{fig:lp_pos} presents the impulse responses for positive shock bins (Bins 1--4). Pre-period estimates are not uniformly flat and are comparable in size to several of post-period estimates, which we flag as a caveat worth bearing in mind when interpreting those results. In the post-period, Bin 2 (positive medium shocks, $n = 140$) and Bin 4 (positive very large shocks, $n = 16$) are negative at every horizon. Bin 2's estimates deepen gradually, from $-0.026$ at $t=0$ to $-0.070$ at $t=+2$ and $-0.161$ at $t=+5$. Bin 4 falls sharply to $-0.212$ at $t=+2$, roughly  three times the magnitude of Bin 2's estimate at the same horizon, before partially recovering to $-0.09$ through $t=+5$. Its confidence intervals are wider, consistent with the smaller sample ($n = 16$). Bins 1 and 3 produce smaller post-period effects, peaking at $-0.058$ and $-0.054$ log points, respectively. Their 95\% confidence bands include zero at every horizon.

\begin{figure}[tbp]
    \centering
    \includegraphics[width=1\textwidth]{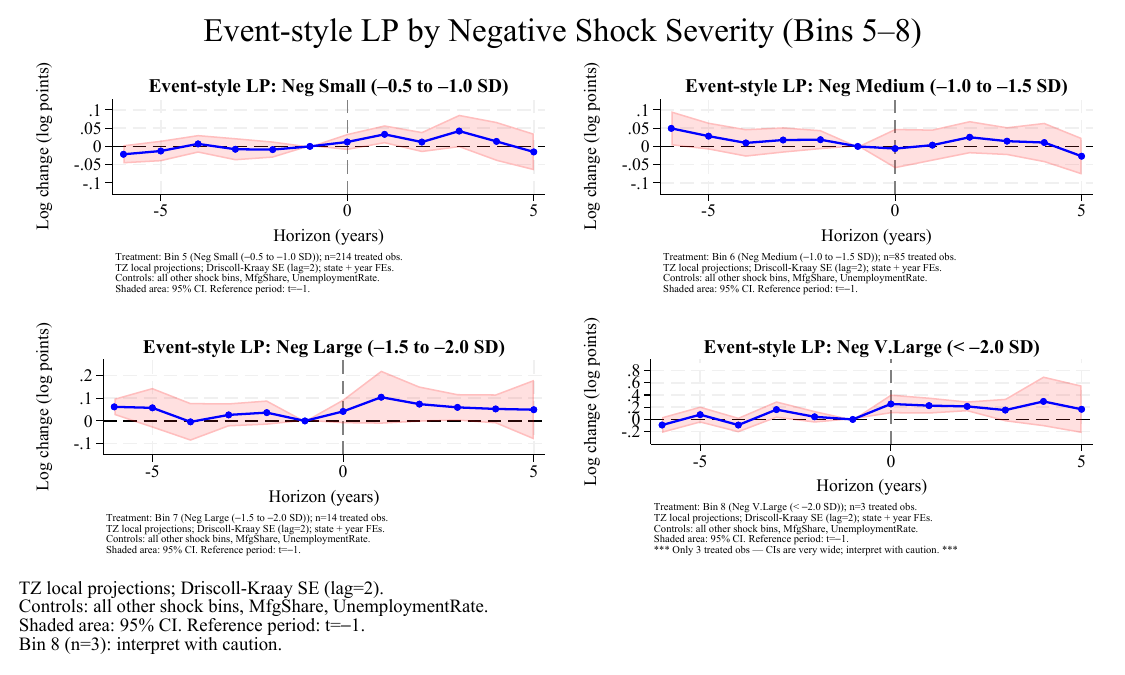}
    \caption{Event-style local projection impulse responses for negative labor share
    shock bins (Bins 5-8). The outcome is the cumulative log change in age-adjusted
    drug mortality relative to $t = -1$. Shaded areas are 95\% confidence intervals
    based on Driscoll--Kraay standard errors (lag 2). All regressions include state
    and year fixed effects, manufacturing employment share, unemployment rate, and
    all other shock bins as nuisance controls. Bin 8 ($n = 3$) should be interpreted
    with considerable caution given the very small number of treated observations.}
    \label{fig:lp_neg}
\end{figure}

Fig.~\ref{fig:lp_neg} presents the impulse responses for negative shock bins (Bins 5--8). They point toward a positive association between negative wage-share shocks and drug-induced mortality that strengthens with shock severity. Bin 5 (negative small shocks, $n = 214$) is mostly near zero but yields one significant result, at $t = +1$ (estimate $= 0.033$, CI $= [0.008, 0.058]$). Bin 6 (negative medium shocks, $n = 85$) is positive in sign at four of six post-shock horizons, peaking at $0.025$ log points at $t = +2$, though none reach significance. Bin 7 (negative large shocks, $n = 14$) shows substantially larger point estimates, peaking at $0.104$ log points at $t = +1$. Moreover, Bin 7 presents a narrowly significant estimate at $t = +3$ (CI $= [0.001, 0.119]$) despite wide intervals elsewhere. Bin 8 (negative very large shocks, $n = 3$) delivers the sharpest pattern. Estimates are significant from $t = 0$ through $t = +2$ ($0.254$, $0.226$, and $0.214$ log points), consistent with an acute mortality response to the most severe labor-share contractions although the fact that we have only three treated observations warrants caution. Effect sizes and precision both rise with shock severity, supporting the model's prediction that deeper wage-share contractions raise drug-induced mortality.

Across Bins 1--8, the dynamic evidence is more asymmetric and uncertain than the static cross-bin comparison suggests. Nonetheless, the group-means pattern and the local projections offer empirical grounding for the model's core mechanism that income distribution dynamics shape substance use behavior and, through this channel, aggregate health outcomes. 

\section{The model}

We consider a closed, one-good economy with a behavioral-health module. The dynamic system consists of three interacting variables. Two are inherited from Goodwin, the employment rate, which plays the role of the prey, and the wage share, which plays the role of the predator. They generate the classical distributive cycle, in which rising employment strengthens workers' bargaining power and compresses profits, while a low wage share restores accumulation and, with it, employment. The third variable is an index of the workforce's composition between agents with and without OUD; it evolves according to a continuous-time discrete-choice mechanism in which the wage share governs the relative attractiveness of the two states. Production is of the Leontief type, workers consume their wages while capitalists invest their profits, so that accumulation is driven by the profit share. Labor productivity grows at an exogenous rate discounted by a damage function that is increasing in OUD prevalence. This damage channel makes the behavioral-health module feeds back into the real economy.

\subsection{Modeling OUD and non-OUD probabilities}

Suppose the workforce ($n$) is constant and divided into two groups, those
with opioid use disorder ($n^{OUD}$) and individuals non-OUD ($n^{NOUD}$), so
that:%
\begin{equation*}
n=n^{OUD}+n^{NOUD}
\end{equation*}%
The fractions of agents in each state evolve according to a continuous-time
version of the discrete-choice approach \citep[e.g.,][]{BrockHommes97}. We
thus write:%
\begin{eqnarray}
\frac{\dot{n}^{OUD}}{n} &=&p^{OUD}\left( U^{OUD}\right) -\frac{n^{OUD}}{n}  \notag \\
&&  \label{dc} \\
\frac{\dot{n}^{NOUD}}{n} &=&p^{NOUD}\left( U^{NOUD}\right) -\frac{n^{NOUD}}{n}  \notag
\end{eqnarray}%
where $p^{OUD}(\cdot)$ and $p^{NOUD}(\cdot)$ are the probabilities of belonging to each group.
They depend on observable factors comprised on $U^{OUD}$ and $U^{NOUD}$.

In line with the discrete-choice literature \citep[for a review see][]{FrankeWesterhoff2017}, we have: 
\begin{eqnarray}
p^{OUD}\left( U^{OUD}\right)  &=&\frac{\exp \left( U^{OUD}\right) }{\exp \left(
U^{OUD}\right) +\exp \left( U^{NOUD}\right) }  \notag \\
&&  \label{pdc} \\
p^{NOUD}\left( U^{NOUD}\right)  &=&\frac{\exp \left( U^{NOUD}\right) }{\exp \left(
U^{OUD}\right) +\exp \left( U^{NOUD}\right) }  \notag
\end{eqnarray}
with both probabilities functions being s-shaped and satisfying $p^{OUD},p^{NOUD} \in (0,1)$ and $p^{OUD}+p^{NOUD}=1$.

We assume that the same variables influence both groups with the opposite sign, i.e. $U^{OUD}=-U^{NOUD}$. Empirical evidence presented in the previous section indicates that reductions in the wage share ($u$) are associated with a higher OUD. On the contrary, a higher size of output going to workers increases the probability of non-OUD. Therefore, we write:%
\begin{equation}
U^{OUD}=-U^{NOUD}=-\gamma u  \label{Udc}
\end{equation}%
where $\gamma >0$ is frequently referred to as the intensity of choice parameter.

Substituting Eq. (\ref{Udc}) into (\ref{pdc}) and the resulting expressions into (\ref{dc}), we have:%
\begin{eqnarray}
\frac{\dot{n}^{OUD}}{n} &=&\frac{\exp \left( -\gamma u\right) }{\exp \left(
-\gamma u\right) +\exp \left( \gamma u\right) }-\frac{n^{OUD}}{n}  \notag \\
&&  \label{dc2} \\
\frac{\dot{n}^{NOUD}}{n} &=&\frac{\exp \left( \gamma u\right) }{\exp \left(
-\gamma u\right) +\exp \left( \gamma u\right) }-\frac{n^{NOUD}}{n}  \notag
\end{eqnarray}

Subtracting the second from the first expression in (\ref{dc2}), we obtain:%
\begin{equation}
\dot{z}=\tanh (-\gamma u)-z \label{z_dynam}
\end{equation} %
where%
\begin{equation*}
z=\frac{n^{OUD}-n^{NOUD}}{n}\in (-1,1)
\end{equation*}%
is an index capturing the composition of the population between those that
OUD and non-OUD agents. When $z=1$, all individuals are in the
first group, while for $z=-1$, nobody abuses opioids. An index $z=0$
indicates the workforce is equally divided between both groups. It goes
without saying that $z$ can be at any point of the spectrum.

\subsection{Introducing the prey}

In Goodwin's (\citeyear{goodwin1967}) original growth cycle model, the employment rate played the role of the ``prey''. Consider an economy that operates with a Leontief production technology:
\begin{equation*}
q=\min \left\{ \frac{k}{v};ena\right\} 
\end{equation*}%
where $q$ stands for output, $k$ is the capital stock, $v=k/q$ is the capital-output ratio, $e=l/n$ is the employment rate, with $l$ as labor demand, and $a=q/l$ corresponds to labor productivity. Provided that $v>0$ is constant and recalling our assumption of a fixed population, from the Leontief dynamic efficiency condition follows that:
\begin{eqnarray}
\frac{\dot{q}}{q} &=&\frac{\dot{k}}{k}  \notag \\
&&  \label{Leontief_eff} \\
\frac{\dot{e}}{e} &=&\frac{\dot{q}}{q}-\frac{\dot{a}}{a}  \notag
\end{eqnarray}%
Substituting the first into the second expression in (\ref{Leontief_eff}),
we have:%
\begin{equation}
\frac{\dot{e}}{e}=\frac{\dot{k}}{k}-\frac{\dot{a}}{a}  \label{e_dynam1}
\end{equation}

All wages are assumed to be consumed, while profits are saved and invested: 
\begin{equation*}
\dot{k}=q-wl=\left( 1-\frac{wl}{q}\right) q=(1-u)q
\end{equation*}%
where $u=wl/q=w/a$ is the share of workers in total output, i.e. the wage share. This implies:%
\begin{equation}
\frac{\dot{k}}{k}=\frac{1-u}{v}  \label{capital_accumulation}
\end{equation}

Labor productivity is assumed to grow at an exogenous rate $\alpha $, but
discounted by the damage effects of drug abuse:
\begin{equation}
\frac{\dot{a}}{a}=[1-D(z)]\alpha   \label{productivity_growth}
\end{equation}%
with $D^{\prime }(\cdot )>0$ indicating that $D(\cdot )$ is an increasing
function in its main argument. A higher share of individuals abusing opioids has a damaging impact on productivity.

Substituting Eqs. (\ref{capital_accumulation}) and (\ref{productivity_growth}%
) into (\ref{e_dynam1}), we obtain the dynamics of the employment rate depending on income distribution and the composition of the population in terms of OUD:%
\begin{equation}
\frac{\dot{e}}{e}=\frac{1-u}{v}-[1-D(z)]\alpha   \label{emp_dynam2}
\end{equation}

The expression above decomposes employment dynamics into two competing forces. The first, $(1-u)/v$, is the classical channel where a higher profit share, $1-u$,
accelerates capital accumulation and pulls employment up. The second, effective
productivity growth $[1-D(z)]\alpha$, enters with a negative sign because productivity
growth is labor-saving. The faster output per worker rises, the fewer workers are
needed to sustain a given path of output. OUD acts on employment through this second term alone. A lower wage share raises OUD prevalence $z$, which increases the damage $D(z)$ and slows effective workers' productivity growth. Since that term is subtracted, the labor-saving drag weakens, and employment grows faster.


\subsection{The predator arrives}

Taking log-derivatives of the wage share definition, $u=w/a$, we have that:%
\begin{equation}
\frac{\dot{u}}{u}=\frac{\dot{w}}{w}-\frac{\dot{a}}{a}
\label{wage_share_dynam1}
\end{equation}
so that the variation rate of our distributive variable will be positive provided that real wages grow faster than labor productivity. The share of income going to workers will fall when productivity outpaces wages.

Suppose the existence of a real-wage Phillips curve:%
\begin{equation}
\frac{\dot{w}}{w}=f\left( e\right)   \label{phillips}
\end{equation}%
with $f^{\prime }(e)>0$ indicating that $f\left( \cdot \right) $ is an
increasing function of the employment rate. For a higher $e$, the bargaining power of workers improves, and they are able to obtain real wage increases.

Substituting Eqs. (\ref{productivity_growth}) and (\ref{phillips}) into Eq. (\ref%
{wage_share_dynam1}), we can write the dynamics of income distribution in
terms of the evolution of employment rates and OUD prevalence:%
\begin{equation}
\frac{\dot{u}}{u}=f\left( e\right) -[1-D(z)]\alpha  \label{wage_share_dynam2}
\end{equation}
The first term, $f(e)$ captures the bargaining position of workers while the second term embodies a channel through which OUD reaches distribution. Higher OUD raises $D(z)$, slows effective productivity growth, and thereby, for a given rate of real wage growth, props up the wage share. The health channel thus enters both laws of motion through the same object, effective productivity growth $[1-D(z)]\alpha$, coupling the composition of the workforce to the employment rate and the wage share at once.

\subsection{The dynamic system}

Our addicted predator--prey model is described by Eqs. (\ref{z_dynam}), (\ref{emp_dynam2}), and (\ref{wage_share_dynam2}), resulting in the following
3-dimensional nonlinear dynamic system: 
\begin{eqnarray}
\dot{e} &=&\left[ \frac{1-u}{v}-\left( 1-D(z)\right) \alpha \right] e  \notag
\\
\dot{u} &=&\left[ f\left( e\right) -\left( 1-D(z)\right) \alpha \right] u
\label{sys1} \\
\dot{z} &=&\tanh (-\gamma u)-z  \notag
\end{eqnarray}%

In steady-state, $\dot{e}=\dot{u}=\dot{z}=0$, giving us the equilibrium conditions:%
\begin{eqnarray}
0 &=&\left[ \frac{1-\bar{u}}{v}-\left( 1-D(\bar{z})\right) \alpha \right] \bar{e}  \notag \\
0 &=&\left[ f\left( \bar{e}\right) -[\left( 1-D(\bar{z})\right) \alpha \right] \bar{u}
\label{equilibrium_conditions1} \\
0 &=&\tanh (-\gamma \bar{u})-\bar{z}  \notag
\end{eqnarray}
Thus, we can state and prove the following Proposition regarding the existence and uniqueness of an internal equilibrium solution:

\begin{proposition} \label{prop 1}
The dynamic system (\ref{sys1}) has a unique economically meaningful equilibrium point $P_1=(\bar{e},\bar{u}, \bar{z})$ defined and given by:
\begin{equation*}
\bar{u}=H^{-1}\left(1-v \alpha \right)
\end{equation*}
\begin{equation*}
\bar{z}=\tanh\left(-\gamma H^{-1}\left(1-v\alpha\right)\right)
\end{equation*}
\begin{equation*}
\bar{e}=f^{-1}\left(\alpha\left[1-D\left(\tanh\left(-\gamma H^{-1}\left(1-v\alpha\right)\right)\right)\right]\right)
\end{equation*}
where
\begin{equation*}
H(u):=u-\alpha vD\left(\tanh\left(-\gamma u\right)\right)
\end{equation*}

\bigskip

\begin{proof}
See Appendix A.1.
\end{proof}

\end{proposition}

The equilibrium is determined recursively, and the order is economically meaningful. Income distribution is pinned down first. Setting $\dot{e}=0$ together with $\dot{z}=0$ yields the equilibrium wage share, which depends on the shape of the discrete-choice mechanism, through $H(\cdot)$, and on the capital-output ratio. Given $\bar{u}$, the composition of the workforce in terms of OUD follows from the switching mechanism. Only then is the employment rate determined, as the value $\bar{e}$ that makes real-wage growth consistent with effective productivity growth. This last step makes the steady-state relationship between health and employment transparent. Since $f(\cdot)$ and $D(\cdot)$ are increasing functions, $\bar{e}$ is a decreasing function of $\bar{z}$. A higher OUD prevalence in steady state is associated with a lower employment rate, consistent with the empirical evidence revisited before.

Next, we turn to the investigation of the local stability properties of $P_1$. Applying the existence part of the Andronov-Hopf bifurcation theorem \citep[p.~479--483]{Gandolfo2009}, we can state and prove the following Proposition:

\begin{proposition} \label{prop 2}
The equilibrium point $P_1=(\bar{e},\bar{u}, \bar{z})$ is locally asymptotically stable in the region of the parameter space defined as:
\begin{equation*}
1-\bar{e}f^{\prime }>0
\end{equation*}%
Moreover, when:
\begin{equation*}
1-\bar{e}f^{\prime }=0
\end{equation*}%
the characteristic equation has a pair of complex conjugate eigenvalues that become purely imaginary, while no other eigenvalues have zero real part. In this case, a Andronov-Hopf bifurcation might occur, and the system admits a family of periodic solutions.
\bigskip

\begin{proof}
See Appendix A.2.
\end{proof}

\end{proposition}

To provide a more concrete view of the properties of the model, we adopt a set of functional forms for the behavioral 
equations of the model. This step allows us to restate the existence and local stability conditions of the equilibrium in terms of 
parameters that later on can be calibrated following empirical studies in the field and well-known time series:
\begin{eqnarray}
D(z)=\theta(1+z) \notag \\ \label{functional_forms}  \\
f(e)=-\beta_0+\beta_1e \nonumber 
\end{eqnarray}
where $\theta>0$ is the sensitivity of damage in labor productivity to OUD; $\beta_0>0$ captures real wage behavior under zero employment; and $\beta_1>0$ is the sensitivity of real wage rate growth to employment. Thus, we can restate Proposition 1 as:

\begin{proposition} \label{prop 3}
The dynamic system $(14)$ admits a unique economically meaningful equilibrium solution that satisfies:
\begin{equation*}
\bar{u}=J^{-1}(1-\alpha v (1-\theta))
\end{equation*}
\begin{equation*}
\bar{z}=\tanh\left(-\gamma J^{-1}(1-\alpha v (1-\theta))\right)
\end{equation*}
\begin{equation*}
\bar{e}=\frac{\beta_0+\alpha \{1-\theta [1+\tanh(-\gamma J^{-1}(1-\alpha v (1-\theta)))]\}}{\beta_1}
\end{equation*}
where
\begin{equation*}
J(u) := u-\alpha v\theta\tanh (-\gamma u)
\end{equation*}

\bigskip

\begin{proof}
See Appendix A.3.
\end{proof}

\end{proposition}

Analogously, we can restate and prove the following Proposition regarding the possibility of a Andronov-Hopf bifurcation to occur:

\begin{proposition} \label{prop 4}
Consider the dynamic system~(\ref{sys1}) under the functional forms~(\ref{functional_forms}). At any parameter configuration satisfying
\begin{equation}
    \beta_0+\alpha \{1-\theta [1+\tanh(-\gamma J^{-1}(1-\alpha v (1-\theta)))]\}=1,
    \label{eq:hopf_condition}
\end{equation}
the Jacobian evaluated at the equilibrium $P_1$ has a pair of purely imaginary eigenvalues. Condition~(\ref{eq:hopf_condition}) thus identifies a candidate Andronov-Hopf bifurcation, at which the system admits a family of periodic solutions.

\bigskip

\begin{proof}
See Appendix A.4.
\end{proof}

\end{proposition}

\section{Numerical simulations}

The analytical results established in Section~3 are now illustrated through a series of numerical experiments. We use the functional forms in~(\ref{functional_forms}) and calibrate the parameters to yield an economically meaningful equilibrium. We stress that this calibration is purely illustrative. Its sole purpose is to bring out the qualitative behavior of the model in a transparent and reproducible way; the values carry no empirical ambition and should not be read as estimates for any specific country, period, or episode. They are reported in Table~\ref{tab:parameters}, and each figure states the parameter values used to generate it.

\begin{table}[tbp]
    \centering
    \caption{Parameter values}
    \label{tab:parameters}
    \begin{tabular}{llll}
        \hline
        Parameter & Value & Description & Motivation \\
        \hline
        $v$        & $6.25$ & Capital-output ratio               & to have $\bar{u}\approx0.75$ \\
        $\alpha$   & $0.05$ & Labor productivity growth rate      & to have $\bar{e}\approx0.6$, $\bar{u}\approx0.75$ \\
        $\beta_0$  & $\in (0.6,0.96)$ & Phillips curve intercept         & to have $\bar{e}\approx0.6$ \\
        $\beta_1$  & $1.67$ & Wage sensitivity to employment      & to have $\bar{e}\approx0.6$ \\
        $\gamma$   & $\in (1.75,1.96)$ & Intensity of choice parameter       & to have $\bar{z}\approx-0.9$ \\
        $\theta$   & $\in (2,3)$ & Productivity damage parameter       & to have $\bar{u}\approx0.75$ \\
        \hline
    \end{tabular}
\end{table}

The model's stylized nature requires loose calibration, since several parameters have to absorb features we deliberately omitted. For example, because there is no depreciation, the capital-output ratio must be set well above its empirical counterpart, at $v=6.25$, to deliver a meaningful wage share. Likewise, with the population held constant, the natural growth rate coincides with the growth rate of labor productivity, which we set at $\alpha=0.05$. We also choose the real-wage Phillips curve parameters on the high side, bringing the system into the region where the Andronov-Hopf bifurcation can be illustrated while delivering a meaningful equilibrium employment rate, $\bar{e}\approx0.6$.

\subsection{Convergence to equilibrium}\label{subsec:convergence}
 
\begin{figure}[tbp]
    \hspace{-2cm}\subfloat[]{\includegraphics[width=1.25\textwidth]{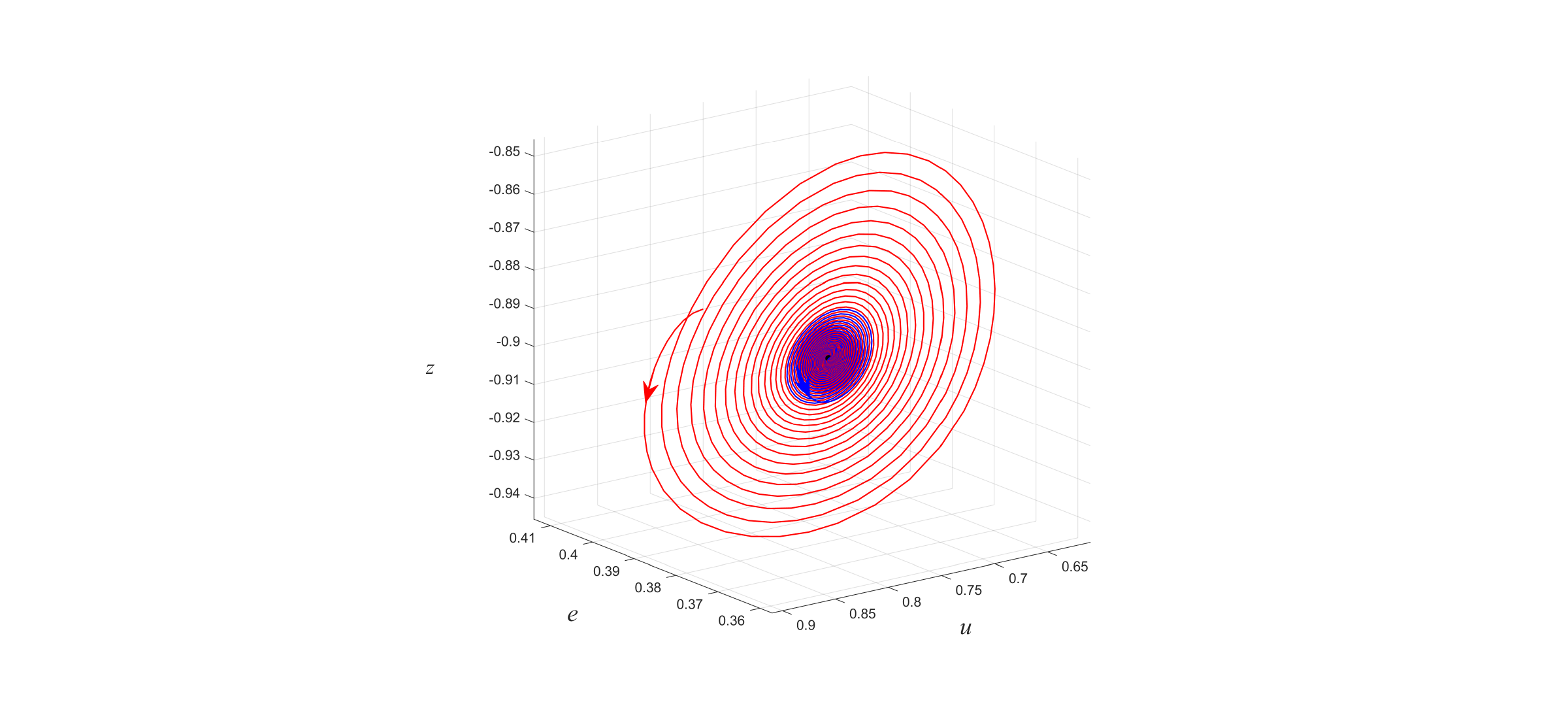}}\\
    \subfloat[]{\includegraphics[width=1\textwidth]{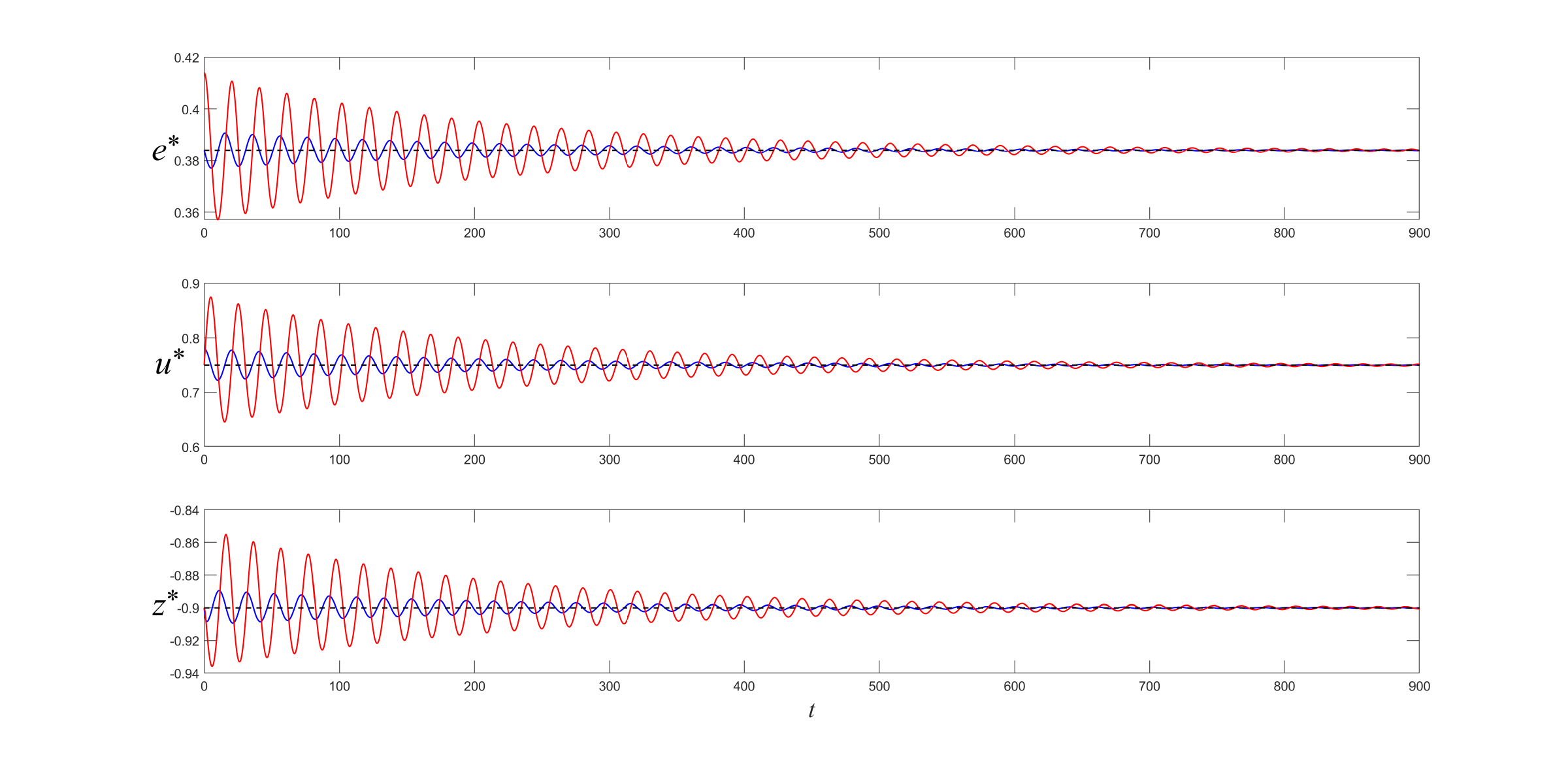}}
    \caption{Convergence to the equilibrium. Parameters: $v = 6.25$, $\alpha = 0.05$, $\beta_0 = 0.6$, $\beta_1 = 1.67$, $\gamma = 1.96$, and $\theta = 2$. We report two trajectories with initial conditions $(\bar{e}, \bar{u}+0.03, \bar{z})$ and $(\bar{e}+0.03, \bar{u}, \bar{z})$, in blue and red, respectively.}
    \label{fig:convergence3d}
\end{figure}

We first show that the model converges to $P_1$ when the local stability conditions in Proposition 2 are satisfied. Fig. \ref{fig:convergence3d} reports two trajectories, in red and blue, converging to our unique internal equilibrium. Panel (a) plots them in the $(e,u,z)$ space, while panel (b) indicates the corresponding time series. An important difference relative to \citet{goodwin1967} is that there, every trajectory is a closed orbit determined entirely by initial conditions, and no path converges to the interior equilibrium. Our three-dimensional extension breaks this conservative structure and can generate asymptotic stability.

\subsection{Coexistence of limit cycles}\label{subsec:limitcycles}
\begin{figure}[tbp]
    \subfloat[]{\includegraphics[width=1\textwidth]{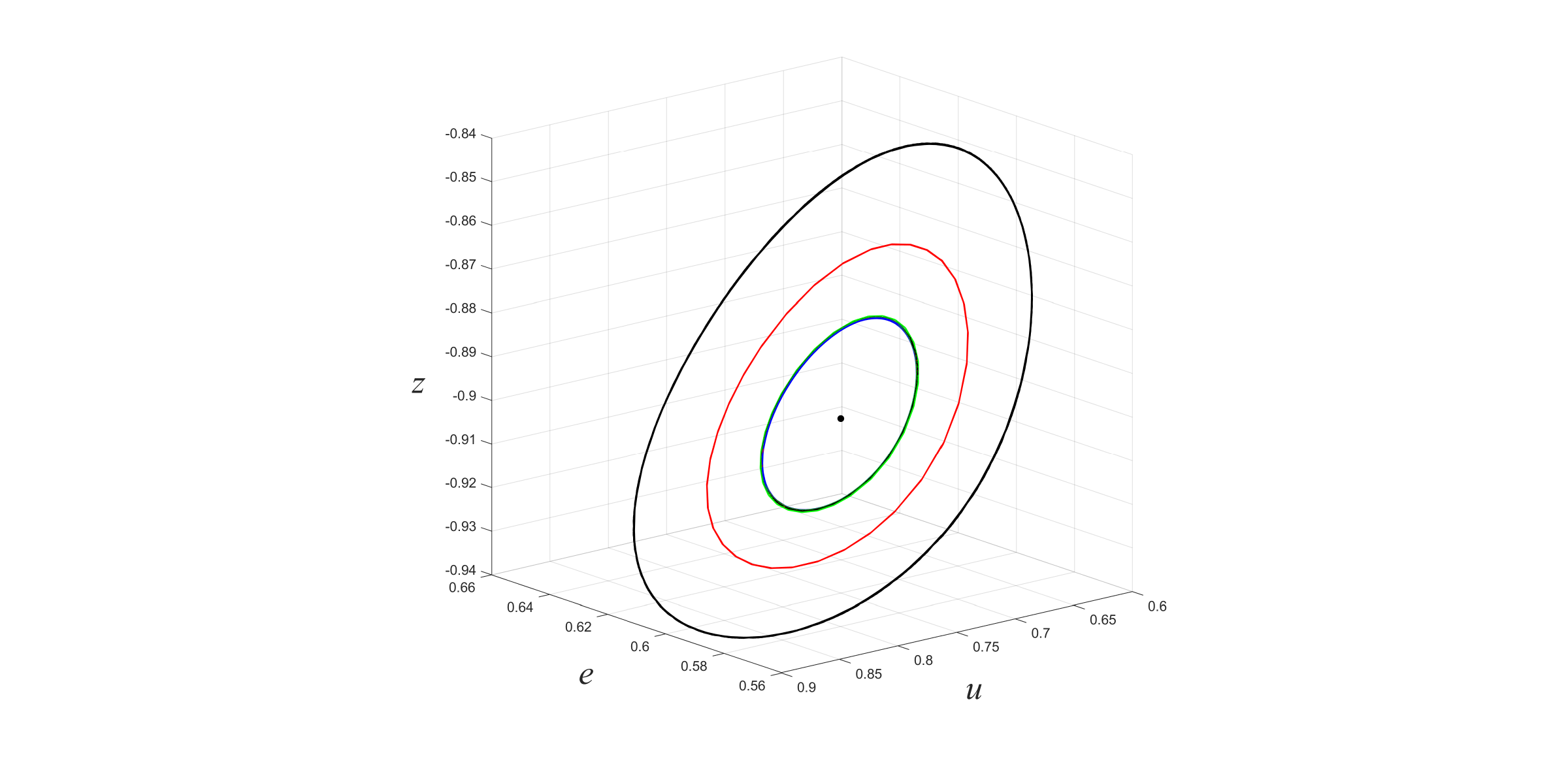}}\\
    \subfloat[]{\includegraphics[width=1\textwidth]{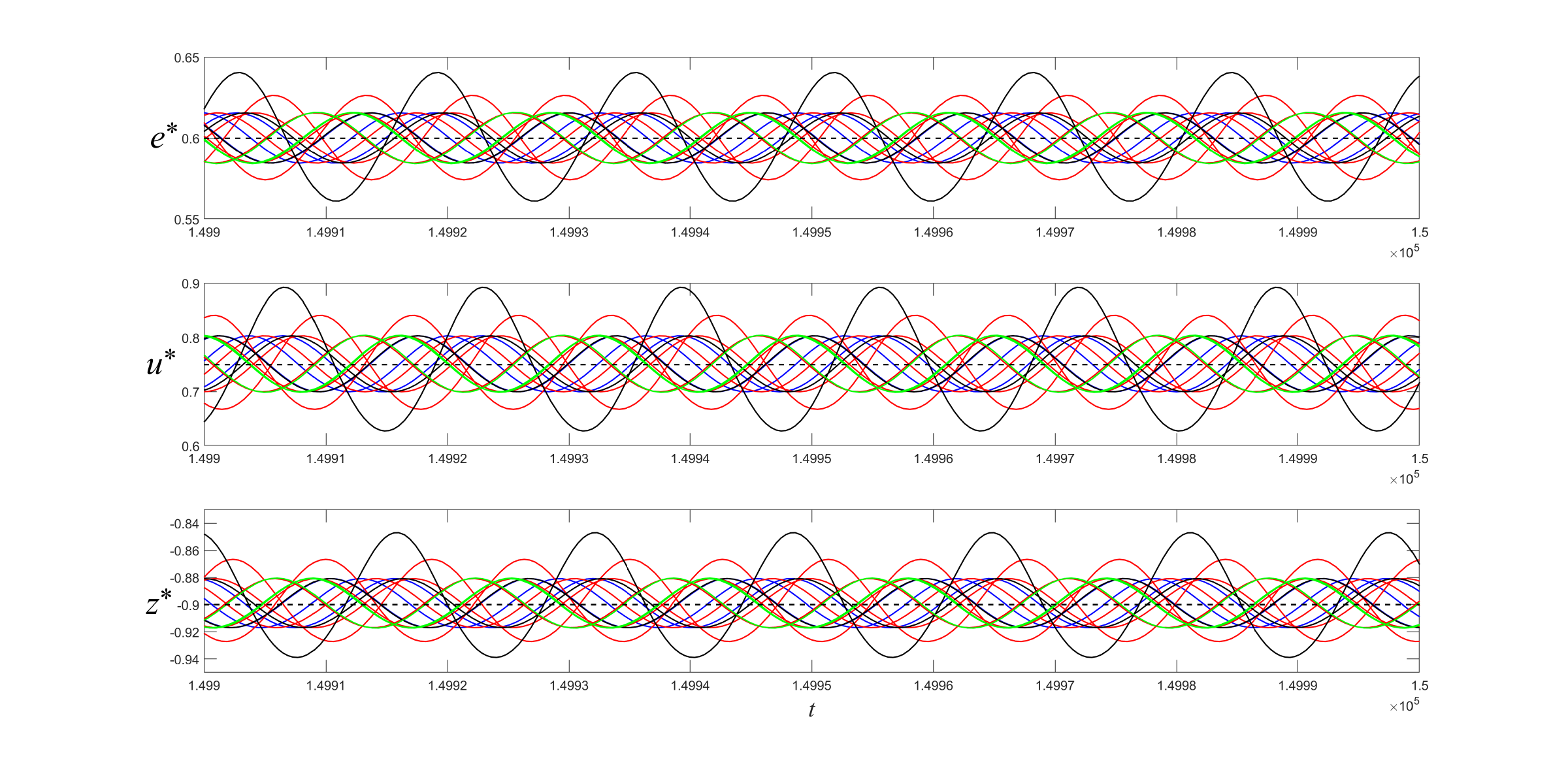}}
    \caption{Coexistence of three limit cycles. Parameters: $v = 6.25$,
    $\alpha = 0.05$, $\beta_0 = 0.96$, $\beta_1 = 1.67$, $\gamma = 1.96$,
    and $\theta = 2$.
    Initial conditions for the blue trajectories are $(\bar{e}+0.01, \bar{u},
    \bar{z})$, $(\bar{e}, \bar{u}+0.01, \bar{z})$, and $(\bar{e}, \bar{u},
    \bar{z}+0.01)$. Initial conditions for the red trajectories are
    $(\bar{e}-0.03, \bar{u}, \bar{z})$, $(\bar{e}, \bar{u}-0.03, \bar{z})$,
    $(\bar{e}, \bar{u}, \bar{z}-0.03)$, $(\bar{e}+0.03, \bar{u}, \bar{z})$,
    $(\bar{e}, \bar{u}+0.03, \bar{z})$, and $(\bar{e}, \bar{u}, \bar{z}+0.03)$.
    Initial conditions for the black trajectories are $(\bar{e}-0.04, \bar{u},
    \bar{z})$, $(\bar{e}, \bar{u}-0.04, \bar{z})$, and $(\bar{e}, \bar{u},
    \bar{z}-0.04)$. Initial conditions for the green trajectories are
    $(\bar{e}-0.01, \bar{u}+0.05, \bar{z})$, $(\bar{e}+0.01, \bar{u}+0.05,
    \bar{z})$, and $(\bar{e}, \bar{u}+0.05, \bar{z}-0.01)$.}
    \label{fig:lc3d}
\end{figure}

The existence part of the Andronov-Hopf bifurcation theorem leaves us in the dark regarding the nature of the bifurcation, which might be sub- or super-critical. Setting $\beta_0 = 0.96$, the condition in Proposition~4 is satisfied. The interior equilibrium $P_1$ loses stability, and the system admits endogenous, persistent oscillations. Fig.~\ref{fig:lc3d} illustrates that our bifurcation is supercritical and, for different initial conditions, the economy settles on one of three distinct limit cycles.

The coexistence of periodic attractors implies that the long-run behavior of the economy is weakly \emph{path dependent}. The amplitude of the orbit the economy converges to depends on initial conditions. Economies that begin with large deviations from $P_1$ lock into high-amplitude oscillations with wider swings in both the wage share and the OUD index, whereas economies that start closer to equilibrium settle into milder, smaller-amplitude cycles. We call this a weak form of path dependence because we do not document chaotic attractors or coexistence across attractor types. Multiplicity is limited to attractors of the same nature.

\subsection{A comparison with the original Goodwin growth cycle}

\begin{figure}[tbp]
    \centering
    \begin{subfigure}[b]{0.48\textwidth}
        \centering
        \includegraphics[width=\textwidth]{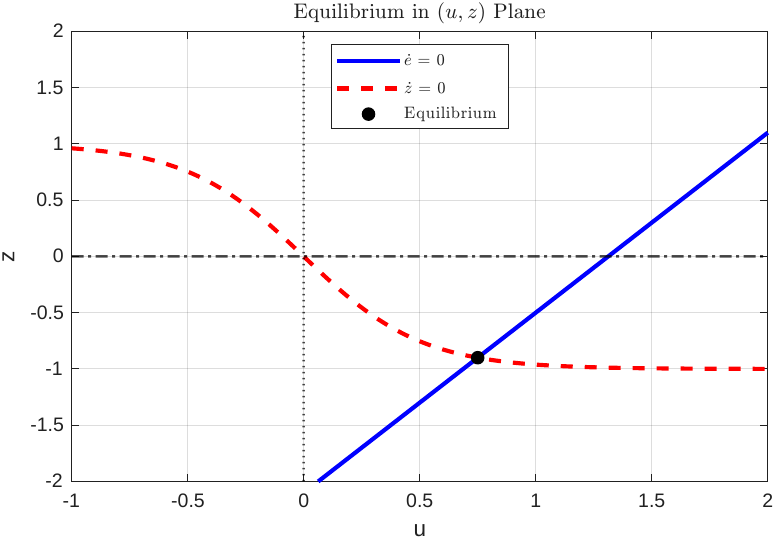}
        \caption{Equilibrium in the $(u, z)$ plane.}
        \label{fig:eq_uz}
    \end{subfigure}
    \hfill
    \begin{subfigure}[b]{0.48\textwidth}
        \centering
        \includegraphics[width=\textwidth]{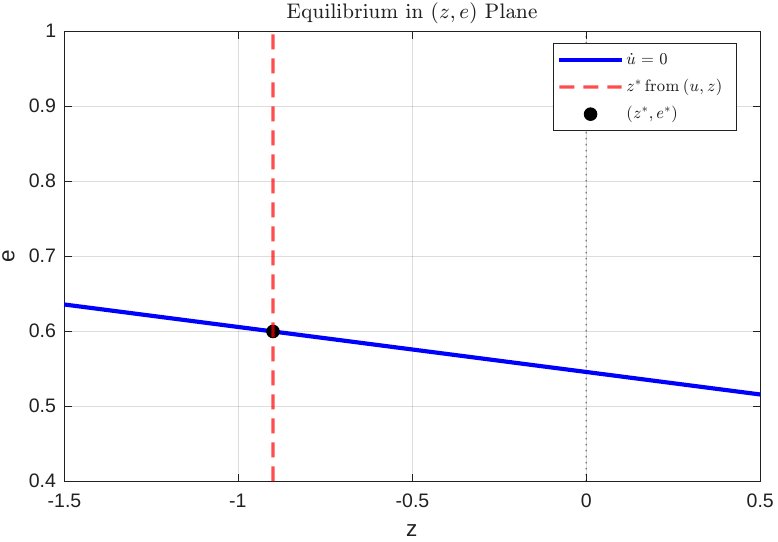}
        \caption{Equilibrium in the $(z, e)$ plane.}
        \label{fig:eq_ze}
    \end{subfigure}
    \caption{Two-step graphical determination of the interior equilibrium
    $P_1 = (\bar{e}, \bar{u}, \bar{z})$. In panel~(a), the solid blue line is
    the $\dot{e}=0$ nullcline and the dashed red curve is the $\dot{z}=0$
    nullcline in the $(u,z)$ plane; their intersection yields
    $(\bar{u}, \bar{z})$. In panel~(b), the solid blue line is the $\dot{u}=0$
    locus in the $(z,e)$ plane; the dashed red vertical line carries forward
    $\bar{z}$ from panel~(a), and their intersection yields
    $\bar{e}$. Parameters: $v = 6.25$, $\alpha = 0.05$, $\beta_0 = 0.96$,
    $\beta_1 = 1.67$, $\gamma = 1.96$, and $\theta = 2$.}
    \label{fig:equilibrium}
\end{figure}

It is instructive to examine how the extended model determines its equilibrium and how it compares with \cite{goodwin1967}. In the canonical two-dimensional growth cycle, the unique interior equilibrium is given by:
\begin{equation}
    \bar{u}_G = 1 - v\alpha, \qquad \bar{e}_G = f^{-1}(\alpha),
    \label{eq:goodwin_eq}
\end{equation}
which can be obtained from (\ref{sys1}) if we set $\theta=\gamma=0$ and $\bar{z}_G$ is determined outside the economic system. In the extended model, OUD modifies both equilibrium values. Substance abuse reduces effective productivity growth, which shifts the equilibrium wage share and employment rate. The OUD index $\bar{z}$, in turn, is endogenously determined by the equilibrium wage share $\bar{u}$ through the discrete-choice mechanism, as illustrated in Fig.~\ref{fig:equilibrium}.

\begin{figure}[tbp]
    \centering
    \subfloat[]{\includegraphics[width=1\textwidth]{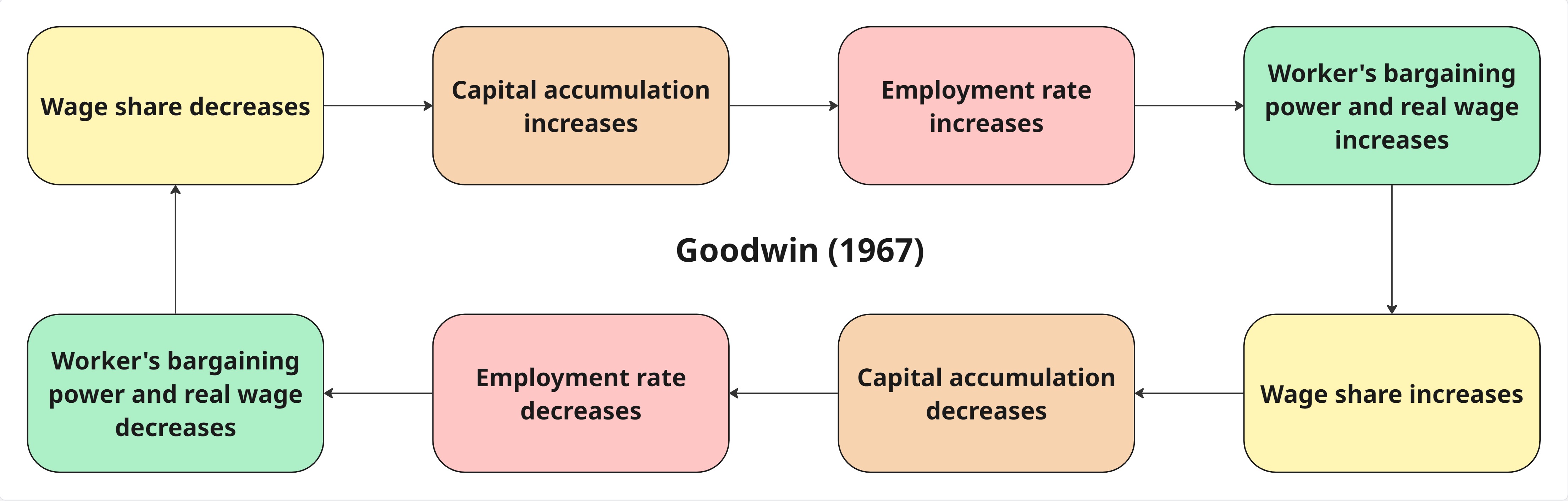}} \bigskip \bigskip  \\
    \subfloat[]{\includegraphics[width=1\textwidth]{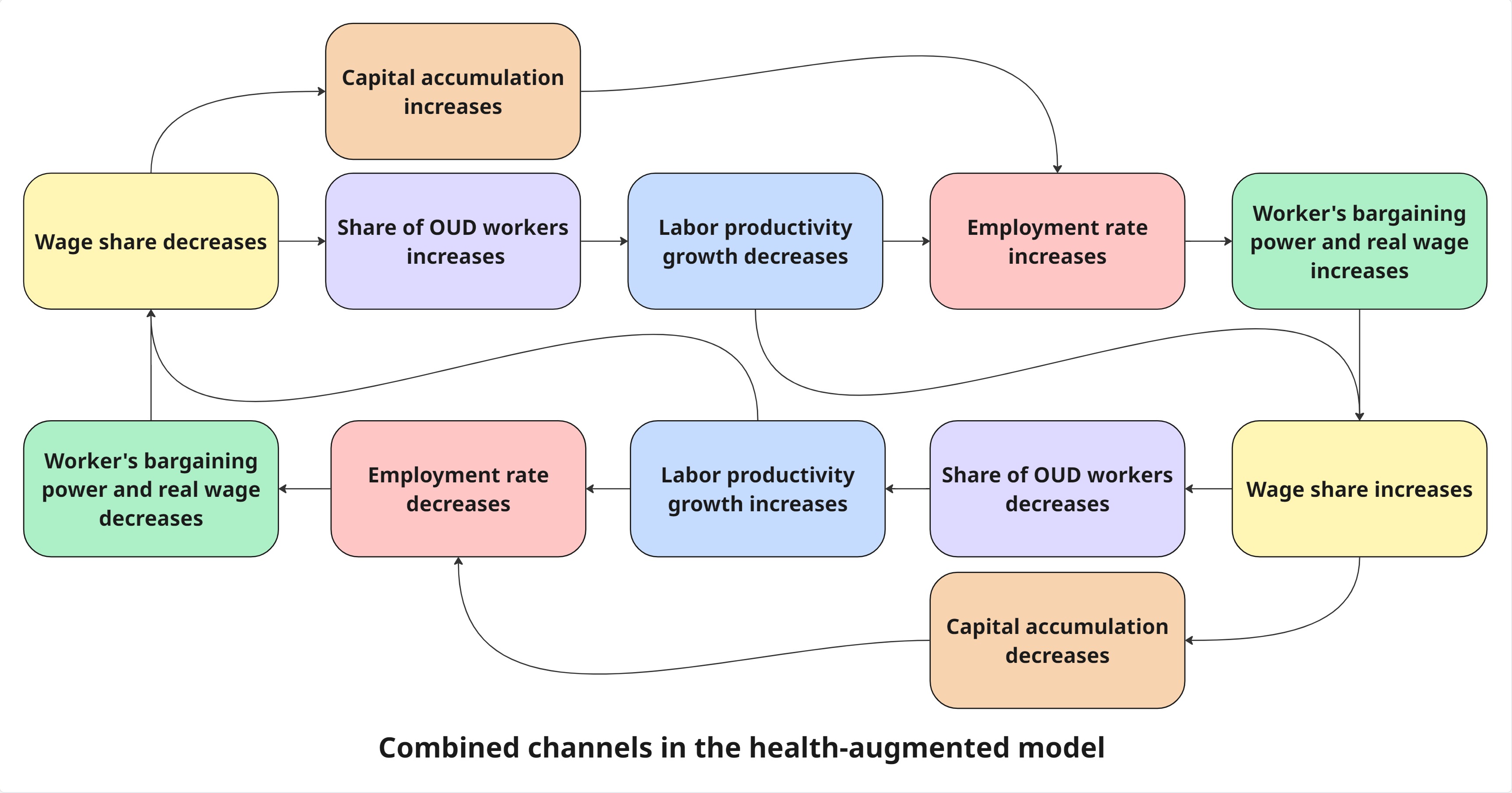}}
    \caption{A comparison diagram between original and health-augmented growth cycles.}
    \label{fig:combined-channels}
\end{figure}

In terms of the interpretation of the cycles, the addicted predator-prey model works as follows. Suppose the wage share decreases. There is an increase in profitability that brings capital accumulation up, improving employment rates. Still, a reduction in the share of income going to workers is also followed by an increase in those with OUD, which damages labor productivity. Lower productivity raises the measured wage share for a given wage and creates the need for more workers for a given level of output, further improving the employment rate. An increasing employment rate improves the bargaining power of workers, leading to higher real wages and bringing the wage share up. This damages profitability, reducing capital accumulation, which ultimately decreases the employment rate as fewer workers are in demand. Because a higher wage share reduces workers with OUD, labor productivity improves, which lowers the measured wage share for a given wage and reduces demand for labor for a given level of output. As employment rates fall, workers' bargaining power also falls, ultimately reducing the wage share. At this point, the cycle restarts. Fig.~\ref{fig:combined-channels}, panels (a) and (b), allow the simple and health-augmented cycles to be compared.

\subsection{Sensitivity of OUD to income
distribution}\label{subsec:gamma}

\begin{figure}[tbp]
    \centering
    \includegraphics[width=0.65\textwidth]{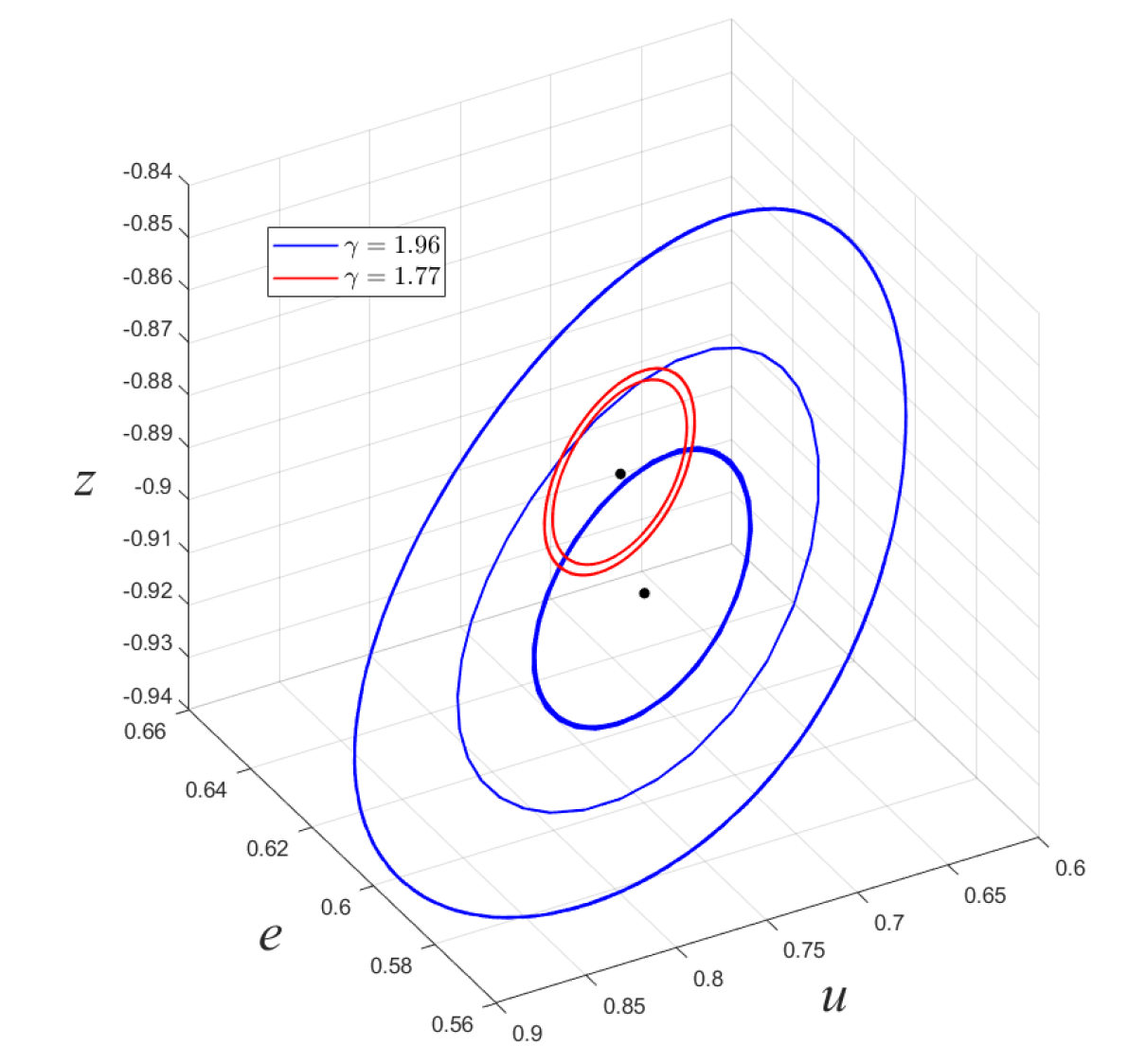}
    \caption{Limit cycle robustness to variations in $\gamma$. Blue orbits: $\gamma = 1.96$,
    with 15 trajectories initialized by displacing $(\bar{e}, \bar{u}, \bar{z})$
    by $\pm 0.01$, $\pm 0.03$, $-0.04$, and mixed perturbations along individual
    state-variable dimensions. Red orbits: $\gamma=
    1.77$, with 15 trajectories initialized by displacing $(\bar{e}, \bar{u}, \bar{z})$ by the same perturbation patterns. Black dots indicate the respective equilibria $P_1$ for each
    parameterization. Parameters: $v = 6.25$, $\alpha = 0.05$,
    $\beta_0 = 0.96$, $\beta_1 = 1.\overline{6}$, and $\theta = 2$.}
    \label{fig:gamma}
\end{figure}

As indicated in the previous subsection, the two critical parameters that differentiate our system from \cite{goodwin1967} are $\gamma$ and $\theta$. We now evaluate the model's sensitivity to variations in the first. The intensity-of-choice parameter governs how strongly the wage share influences workers' transition probabilities between OUD and non-OUD states. Fig.~\ref{fig:gamma} compares limit cycles for $\gamma=1.77$, in red, and $\gamma = 1.96$, in blue. A lower value of this parameter shifts $P_1$ to a higher OUD, a higher wage share, and essentially the same employment rate.

The causal chain runs as follows. A smaller $\gamma$ flattens the $\dot{z}=0$ locus $\bar{z} = \tanh(-\gamma\bar{u})$, so that for any given wage share workers respond less strongly to income in their
substance-use decisions and $\bar{z}$ becomes less negative. The resulting rise in OUD prevalence amplifies the productivity damage $D(\bar{z}) =\theta(1+\bar{z})$, which lowers effective productivity growth $[1-D(\bar{z})]\alpha$. To restore the $\dot{e}=0$ condition $(1-\bar{u})/v = [1-D(\bar{z})]\alpha$, the economy requires a higher wage share $\bar{u}$. The near-unchanged employment rate $\bar{e}$ follows directly from the $\dot{u}=0$ condition, which pins down $\bar{e}$ independently of $\bar{u}$.
 
Second, a reduction in $\gamma$ compresses the amplitude of limit cycles. The red orbits are substantially smaller than the blue orbits along every dimension of the state space. This compression reflects a weaker health-channel entry point, as Fig.~\ref{fig:combined-channels} makes explicit. Moreover, we also document the coexistence of two instead of three orbits. Under high $\gamma$, a downturn in the wage share triggers a pronounced increase in OUD prevalence, which discounts productivity growth through the damage function $D(z)$ and further raises employment; the resulting tighter labor market then feeds through the Phillips curve into a sharper rebound in the wage share, a destabilizing spiral that sustains wide oscillations.
 
From a policy perspective, $\gamma$ reflects institutional and social conditions behind the link between economic stress and substance abuse. We argue that factors that effectively reduce $\gamma$ include the generosity of social safety nets, universal access to healthcare and mental health services, housing security, and the availability of non-pharmacological coping mechanisms. When these buffers are in place, workers' substance-use behavior becomes less sensitive to cyclical income fluctuations, weakening the amplification channel between income distribution and health. Under this interpretation, our model suggests that social policy has a macroeconomic stabilizing role beyond its direct health benefits. It can potentially reduce both the amplitude of the growth cycle and the endogenous variation in workforce health composition over the business cycle.
 
\subsection{Responsiveness of labor productivity to
OUD}\label{subsec:theta}
 
\begin{figure}[tbp]
    \centering
    \includegraphics[width=0.65\textwidth]{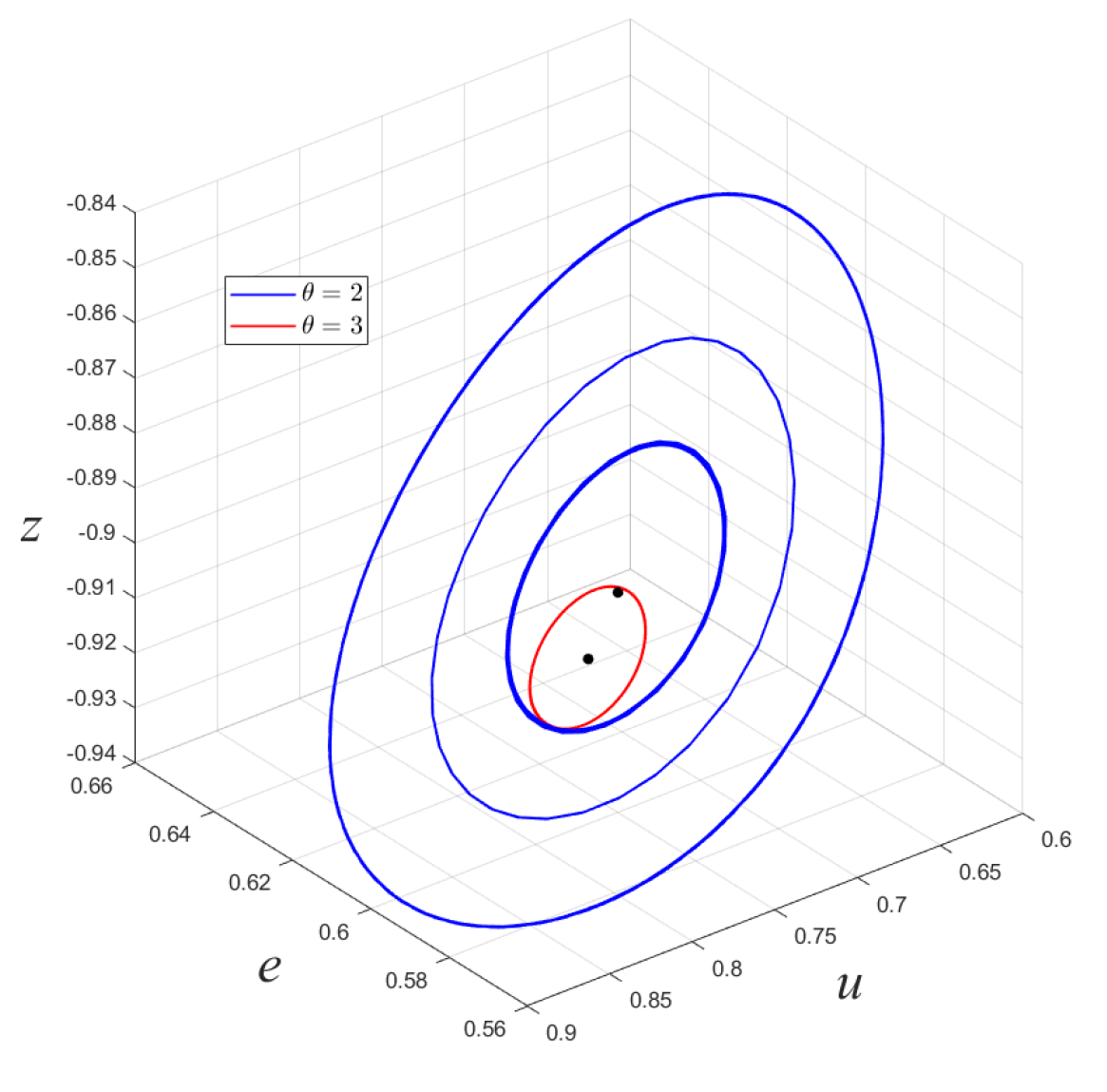}
    \caption{Limit cycle robustness to variations in $\theta$. Blue orbits: $\theta = 2$,
    with 15 trajectories initialized by displacing $(\bar{e}, \bar{u}, \bar{z})$
    by $\pm 0.01$, $\pm 0.03$, $-0.04$, and mixed perturbations along individual
    state-variable dimensions. Red orbit: $\theta = 3$, with 15 trajectories
    initialized by displacing equilibrium $(\bar{e},
    \bar{u}, \bar{z})$ by the same perturbation patterns. Black dots indicate
    the respective equilibria $P_1$ for each parameterization. Parameters:
    $v = 6.25$, $\alpha = 0.05$, $\beta_0 = 0.96$, $\beta_1 = 1.\overline{6}$,
    and $\gamma = 1.96$.}
    \label{fig:theta}
\end{figure}

Parameter $\theta$ determines the magnitude of the productivity damage that substance abuse inflicts on the economy. A higher $\theta$ therefore corresponds to an economy in which OUD workers impose a heavier toll on aggregate productivity. This could be the result, for instance, of a higher concentration of employment in physically demanding occupations susceptible to opioid-related impairment \citep{Lundstrom2025}, less access to workplace treatment accommodation, or more severe forms of addiction with greater presenteeism and absenteeism costs. Fig.~\ref{fig:theta} compares limit cycles for $\theta = 2$, in blue, and $\theta = 3$, in red.

Convergence to each orbit is slow, and the diagram disregards those transitory interactions. With a higher value of $\theta$, we again find essentially no change in equilibrium employment rates. However, the wage share and OUD prevalence move in the opposite direction, with a higher $\theta$ raising the wage share while lowering OUD prevalence. The mechanism follows from Proposition~3. A higher $\theta$ amplifies the productivity cost of any given level of OUD prevalence, lowering effective productivity growth $[1-D(\bar{z})]\alpha$. To restore the $\dot{e}=0$ condition $(1-\bar{u})/v = [1-D(\bar{z})]\alpha$, the economy must sustain a higher wage share $\bar{u}$. Since $\bar{z} = \tanh(-\gamma\bar{u})$, the higher $\bar{u}$ in turn pushes $\bar{z}$ slightly more negative, reflecting the self-correcting tendency toward a more non-OUD workforce when health costs are severe. The near-unchanged employment rate $\bar{e}$ again follows from the $\dot{u}=0$ condition, which is independent of $\bar{u}$.
 
We also find that a stronger productivity-damage parameter compresses the amplitude of the limit cycles. The red orbit is substantially smaller than even the innermost blue orbit. Because effective productivity growth enters $\dot{u}$ negatively, a higher $\theta$ pushes the wage share up immediately, for a given real wage. Let us call this effect the \textit{mechanical channel}. Over the cycle, a falling wage share raises capital accumulation and OUD, and both raise the employment rate; the wage share recovers only later, once the tighter labor market has fed through the Phillips curve. The \textit{mechanical channel}, by contrast, lifts the wage share at once, so that $u$ turns around earlier than the employment-mediated channel alone would deliver. Capital accumulation and OUD therefore begin to reverse before the expansion has run its full course, and each swing is cut shorter. This does not overturn the cycle. The employment channel remains the amplifying force that sustains the oscillation. In fact, the orbit persists at $\theta=3$, so the direct channel operates as a smoother rather than a dominant force. A higher $\theta$ adds damping at the margin, enough to shrink the limit cycle but not to eliminate it.


 
This result carries a sharp and non-trivial policy implication. Consider two economies that differ only in $\theta$. The high-$\theta$ economy will exhibit a more drug-clean workforce at equilibrium and smaller cyclical swings in employment and wages, but it will suffer larger productivity losses per OUD worker. The low-$\theta$ economy tolerates a higher equilibrium share of OUD workers and wider macroeconomic fluctuations, but with lower damage per affected individual. Interventions that
reduce $\theta$, such as workplace treatment programs, harm-reduction policies, or occupational redesign that lowers the productivity impact of OUD, may therefore inadvertently \emph{widen} the growth cycle while improving individual health outcomes. A comprehensive policy assessment must weigh these macroeconomic dynamic effects alongside the direct welfare benefits of reducing the severity of
substance abuse.
 
The contrasting effects of $\gamma$ and $\theta$ on the orbit structure reflect their distinct positions in Fig.~\ref{fig:combined-channels}. A reduction in $\gamma$ weakens the \emph{entry point} of the health channel, i.e., the arrow connecting income distribution to substance abuse. An increase in $\theta$, by contrast, strengthens an \emph{interior link} of the health channel, i.e., the arrow connecting OUD prevalence to productivity growth. While $\theta$ governs the strength of the productivity transmission once health behavior has changed, $\gamma$ rules the width of the gate between economic stress and health behavior.

\section{Final considerations}

This paper has extended \citet{goodwin1967}'s predator-prey growth-cycle model to incorporate the endogenous link between income distribution and substance abuse, and the feedback from health behavior to aggregate labor productivity. The proposed framework introduces a population composition index that evolves according to a continuous-time discrete-choice mechanism where a decline in the wage share is associated with a higher probability of opioid use disorder among workers. This index enters the economy through a productivity damage function, transforming what is typically treated as a static health externality into a dynamic macroeconomic channel that interacts with the investment and wage dynamics over the growth cycle.

We provide empirical evidence supporting the model's key assumption, while also highlighting its limits. The static cross-bin relationship between labor share shock severity and age-adjusted drug mortality is consistent with the income-compression channel assumed in the model. The dynamic local projections further suggest that positive labor share gains, particularly at moderate severity, reduce drug-induced mortality over horizons of one to four years. The response to negative shocks is less precisely estimated at severe levels, where treated observations are scarce, and near-zero at mild levels. Whether this asymmetry reflects a genuine nonlinearity in the health-behavior response or an identification limitation remains open.

Applying the existence part of the Andronov-Hopf bifurcation theorem, we show that the system admits a family of periodic orbits. Numerical simulations revealed the coexistence of up to three limit cycles, providing a weak representation of path dependence. Economies starting from larger employment imbalances are drawn into high-amplitude cycles with wider swings in wages, employment, and workforce health composition. Our numerical experiments further yield two contrasting policy lessons. On the one hand, reducing the intensity of choice parameter, governing how strongly substance-use behavior responds to income fluctuations, reduces the macroeconomic instability generated by the health channel. Thus, policies that weaken the link between economic stress and OUD, such as expanding social safety nets, universal access to mental health and substance-use treatment, and housing security programs, carry a macroeconomic stabilizing dividend beyond their direct health benefits. On the other hand, reducing the productivity damage per OUD worker does not unambiguously stabilize the cycle. While it lowers the level of productivity loss at the equilibrium, it also weakens a self-correcting feedback loop that runs from OUD-induced productivity damage back through the wage share. Harm-reduction policies or workplace accommodation programs that reduce the on-the-job productivity cost of OUD should therefore be evaluated with awareness of this potential macroeconomic dynamic effect.

Several extensions of the present framework merit future investigation. First, the model abstracts from the financial sector, treating investment as entirely profit-funded and ignoring debt dynamics. Incorporating Minsky's financial instability hypothesis, so that firms' debt accumulation amplifies or dampens the growth cycle, would allow the model to address how financial fragility interacts with the health-behavior channel and whether periods of high indebtedness magnify the macroeconomic consequences of labor income compression. Second, the workforce is held constant, and labor force participation is treated as separable from the productivity mechanism. A richer treatment would allow OUD prevalence to directly reduce the effective supply of labor, not only its productivity, amplifying employment dynamics during downturns in the spirit of \citet{Krueger2017} and \citet{GreenwoodGunerKopecky2022}. Third,
the analysis is confined to an aggregate closed economy. A multi-region or open-economy extension could address how the geographic concentration of negative labor share shocks documented in the data interacts with the aggregate cycle. These extensions are left for future research.

\newpage

\appendix

\numberwithin{equation}{section}
\numberwithin{figure}{section}

\section{Mathematical Appendix} 

\subsection{Proof of Proposition 1} 

Recall the equilibrium conditions:%
\begin{eqnarray*}
0 &=&\left[ \frac{1-u}{v}-\left( 1-D(z)\right) \alpha \right] e \\
0 &=&\left[ f\left( e\right) -\left( 1-D(z)\right) \alpha \right] u \\
0 &=&\tanh (-\gamma u)-z
\end{eqnarray*}%
An economically meaningful solution requires $e\neq 0$ and $u\neq 0$.
Therefore, it follows:%
\begin{eqnarray}
0 &=&\frac{1-u}{v}-\left( 1-D(z)\right) \alpha   \label{eqcond1} \\
0 &=&f\left( e\right) - \left( 1-D(z)\right) \alpha   \label{eqcond2} \\
z &=&\tanh (-\gamma u)  \label{eqcond3}
\end{eqnarray}

From Eq. (\ref{eqcond1}), we have that:
\begin{align}
1-u& =v\alpha -vD(\tanh (-\gamma u))\alpha   \notag \\
u-vD(\tanh (-\gamma u))\alpha & =1-v\alpha   \label{eqcond1_2}
\end{align}%
Define the auxiliary function:%
\begin{equation*}
H(u)=u-vD(\tanh (-\gamma u))\alpha 
\end{equation*}%
Substituting in Eq. (\ref{eqcond1_2}), we obtain:%
\begin{equation*}
H(u)=1-v\alpha 
\end{equation*}
from which it follows that 
\begin{equation*}
\bar{u}=H^{-1}(1-v\alpha)
\end{equation*}%
provided that the inverse of $H\left( \cdot \right) $ exists.

Substituting $\bar{u}$ into Eq. (\ref{eqcond3}), the equilibrium value of the OUD index is given by:%
\begin{equation*}
\bar{z}=\tanh (-\gamma H^{-1}(1-v\alpha ))
\end{equation*}%
Moreover, substituting $\bar{u}$ and $\bar{z}$ into (\ref{eqcond2}), we have:%
\begin{align*}
f(e)& =\alpha \lbrack 1-D(\tanh(-\gamma u))] \\
\bar{e}& =f^{-1}(\alpha \lbrack 1-D(\tanh(-\gamma u))]) \\
& =f^{-1}(\alpha \lbrack 1-D(\tanh (-\gamma H^{-1}(1-v\alpha )))])
\end{align*}
Still, the demonstration crucially depends on the existence of $H(\cdot)$ inverse. To prove this, we need to show whether $H\left( \cdot \right) $
is bijective. That is, injective, each input maps to a unique output, and
surjective, every possible output is mapped from some input. \medskip

\textbf{Proving injectivity:}
If a function $g$ is differentiable on an interval $I$, and its derivative $g' (x)$ is always positive or always negative on that interval, then $g$ is strictly monotonic, which implies injectivity. In our
case: 
\begin{align*}
H(u)& =u-\alpha vD(\tanh (-\gamma u)) \\
\frac{\partial H}{\partial u}& =1+\alpha v\gamma D^{\prime}(1-\tanh^2(-\gamma u)))>0
\end{align*}%
Thus, $H(u)$ is strictly increasing and injective.

\textbf{Proving surjectivity:}
Using the Intermediate Value Theorem (IVT), if a function $g$ is continuous
on an interval $[a,b]$, and $g(a)<y<g(b)$ (or vice versa), then there exists
some $c\in (a,b)$ such that: 
\begin{equation*}
g(c)=y
\end{equation*}%
In our case, it requires to check limits at infinity: 
\begin{align*}
\lim_{u\rightarrow 0}H(u)& =\lim_{u\rightarrow 0}u-v\alpha D(\tanh (-\gamma
u))=-vD(0)\alpha  \\
\lim_{u\rightarrow 1}H(u)& =\lim_{u\rightarrow 1}u-v\alpha D(\tanh (-\gamma
u))=1-v\alpha D(\tanh (-\gamma ))
\end{align*}%
Thus, by the IVT, for every $y\in \lbrack -vD(0)$,$1-v\alpha D(\tanh(-\gamma
))]$, there exists $u\in \lbrack 0,1]$ such that $H(u)=y$. Therefore, $H(u)$
is surjective onto the interval $[-vD(0)$,$1-v\alpha D(\tanh(-\gamma ))]$.

\subsection{Proof of Proposition 2}
The Jacobian matrix evaluated at $P_1$ is given by:%
\begin{equation*}
J=%
\begin{bmatrix}
j_{11} & j_{12} & j_{13} \\ 
j_{21} & j_{22} & j_{23} \\ 
j_{31} & j_{32} & j_{33}%
\end{bmatrix}%
\end{equation*}%
where%
\begin{align*}
j_{11}&=\left. \frac{\partial \dot{e}}{\partial e}\right\vert _{\bar{e},\bar{u%
},\bar{z}}=0 \qquad \qquad j_{12}=\left. \frac{\partial \dot{e}}{\partial u}%
\right\vert _{\bar{e},\bar{u},\bar{z}}=-\frac{\bar{e}}{v} \qquad \qquad j_{13}=%
\left. \frac{\partial \dot{e}}{\partial z}\right\vert _{\bar{e},\bar{u},\bar{%
z}}=\alpha \bar{e}D^{\prime} \\
j_{21}&=\left. \frac{\partial \dot{u}}{\partial e}\right\vert _{\bar{e},\bar{u%
},\bar{z}}=\bar{u}f^{\prime } \qquad \qquad j_{22}=\left. \frac{\partial \dot{u}}{%
\partial u}\right\vert _{\bar{e},\bar{u},\bar{z}}=0 \qquad \qquad j_{23}=\left. 
\frac{\partial \dot{u}}{\partial z}\right\vert _{\bar{e},\bar{u},\bar{z}%
}=\alpha \bar{u}D^{\prime } \\
j_{31}&=\left. \frac{\partial \dot{z}}{\partial e}\right\vert _{\bar{e},\bar{u%
},\bar{z}}=0 \qquad \qquad j_{32}=\left. \frac{\partial \dot{z}}{\partial u}%
\right\vert _{\bar{e},\bar{u},\bar{z}}=-\gamma \left[ 1-\tanh ^{2}(-\gamma \bar{u})%
\right] \qquad \qquad j_{33}=\left. \frac{\partial \dot{z}}{\partial z}%
\right\vert _{\bar{e},\bar{u},\bar{z}}=-1
\end{align*}
Thus, the characteristic equation can be written as 
\[
\lambda^3 + a_1 \lambda^2 + a_2 \lambda + a_3 = 0
\]
where the coefficients are given by 
\begin{equation*}
\text{(i)\ \ \ }a_{1}=-\text{tr}J=-(0+0-1)=1>0
\end{equation*}%
\noindent 
\begin{equation*}
\text{(ii)}\ \ a_{2}=%
\begin{vmatrix}
0 & \alpha \bar{u}D^{\prime } \\ 
-\gamma \left[ 1-\tanh ^{2}(-\gamma \bar{u})\right]  & -1%
\end{vmatrix}%
+%
\begin{vmatrix}
0 & -\frac{\bar{e}}{v} \\ 
\bar{u}f^{\prime } & 0%
\end{vmatrix}%
=\alpha \bar{u}\gamma D^{^{\prime }}\left[ 1-\tanh ^{2}(-\gamma \bar{u})\right] +\bar{u}f^{\prime }\frac{\bar{e}}{v}>0
\end{equation*}%
\noindent 
\begin{eqnarray*}
\text{(iii) \ \ }a_{3} &=&-\det J=\bar{u}f^{\prime }\left\{ \frac{\bar{e}}{v}%
+\alpha \bar{e}\gamma D^{\prime }\left[ 1-\tanh ^{2}(-\gamma \bar{u})\right] \right\}  \\
&=&\alpha \bar{e}\bar{u}\gamma D^{\prime }f^{\prime }\left[ 1-\tanh ^{2}(-\gamma \bar{u})\right]
+\bar{u}f^{\prime }\frac{\bar{e}}{v}>0
\end{eqnarray*}%
\noindent 
\begin{eqnarray}
\text{(iv) \ \ }a_{1}a_{2}-a_{3} &=&\alpha \bar{u}\gamma D^{\prime }\left[ 1-\tanh
^{2}(-\gamma \bar{u})\right] +\bar{u}f^{\prime }\frac{\bar{e}}{v}-\alpha \bar{e}\bar{u}\gamma D^{\prime} f^{\prime
}\left[ 1-\tanh ^{2}(-\gamma \bar{u})\right] -\bar{u}f^{\prime }\frac{\bar{e}}{v}
\nonumber \\
&=&\alpha \bar{u}\gamma D^{\prime}\left[ 1-\tanh ^{2}(-\gamma \bar{u})\right] (1-\bar{e}f^{\prime })\gtreqqless 0 \label{critical_condition}
\end{eqnarray}
The necessary and sufficient conditions for the local stability of a given equilibrium point is that all roots of the characteristic equation have negative real parts. From the Routh-Hurwitz conditions, this requires:
\[a_1 > 0\]
\[a_2 > 0\]
\[a_3 > 0\]
\[a_1a_2 - a_3 >0\]
Thus, the critical condition is the last one, and the internal equilibrium $P_{1}$ is locally stable provided that: 
\begin{equation*}
1-\bar{e}f^{\prime }>0
\end{equation*}%
On the other hand, a Andronov-Hopf bifurcation can occur for values of the parameters such that $a_1a_2-a_3=0$. That is when:
\begin{equation*}
\bar{e}f^{\prime }=1
\end{equation*}%

\subsection{Proof of Proposition 3}
Disregarding the trivial case in which $\bar{u}=\bar{z}=\bar{e}=0$, the equilibrium of the dynamic system (\ref{sys1}) imply:
\begin{eqnarray}
0 &=&\frac{1-u}{v}-\left( 1-D(z)\right) \alpha \notag \\
0 &=&f\left( e\right) - \left( 1-D(z)\right) \alpha \notag \\
z &=&\tanh (-\gamma u) \notag
\end{eqnarray}
Using the functional forms in (\ref{functional_forms}), we can rewrite them as:
\begin{eqnarray}
[1-\theta(1+z)]\alpha &=&\frac{1-u}{v} \label{eqcond1_new} \\
\left[1-\theta(1+z)\right]\alpha &=&-\beta_0+\beta_1 e \label{eqcond2_new} \\
z &=&\tanh (-\gamma u) \label{eqcond3_new} 
\end{eqnarray}
By substituting Eq. (\ref{eqcond3_new}) into Eq. (\ref{eqcond1_new}), we have that:
\begin{align}
1-u& =\alpha v[1-\theta(1+\tanh(-\gamma u))]   \notag \\
u-\alpha v\theta \tanh (-\gamma u))& =1-\alpha v+\alpha v\theta   \label{eqcond1_new2}
\end{align}%
Define the auxiliary function:%
\begin{equation*}
J(u) := u-\alpha v\theta\tanh (-\gamma u) 
\end{equation*}%
Substituting it into Eq. (\ref{eqcond1_new2}), we obtain:%
\begin{equation*}
J(u)=1-\alpha v (1-\theta)
\end{equation*}
from which it follows that 
\begin{equation*}
\bar{u}=J^{-1}(1-\alpha v (1-\theta))
\end{equation*}%
provided that the inverse of $J\left( \cdot \right) $ exists.

Substituting $\bar{u}$ into Eq. (\ref{eqcond3_new}), it follows that:%
\begin{equation*}
\bar{z}=\tanh (-\gamma J^{-1}(1-\alpha v (1-\theta)))
\end{equation*}%
Moreover, substituting $\bar{z}$ into (\ref{eqcond2_new}), we have:%
\begin{align}
\beta_1 \bar{e}& =\beta_0+[1-\theta(1+\bar{z})]\alpha \nonumber \\
\bar{e}& =\frac{\beta_0+\alpha \{1-\theta [1+\tanh(-\gamma J^{-1}(1-\alpha v (1-\theta)))]\}}{\beta_1} \label{e_equilibrium} 
\end{align}
Still, the demonstration crucially depends on the existence of $J(\cdot)$ inverse. To prove this, we need to show whether $J\left( \cdot \right) $
is bijective. That is, injective, each input maps to a unique output, and surjective, every possible output is mapped from some input. \medskip

\textbf{Proving injectivity:}
If a function $g$ is differentiable on an interval $I$, and its derivative $g' (x)$ is always positive or always negative on that interval, then $g$ is strictly monotonic, which implies injectivity. In our
case: 
\begin{align*}
J(u)& =u-\alpha v\theta\tanh (-\gamma u) \\
\frac{\partial J}{\partial u}& =1+\alpha v\gamma\theta (1-\tanh^2(-\gamma u))>0
\end{align*}%
Thus, $J(u)$ is strictly increasing given parameters $\alpha, v, \gamma, \theta$ are positive and injective.

\textbf{Proving surjectivity:}
Using the IVT again, if a function $g$ is continuous on an interval $[a,b]$, and $g(a)<y<g(b)$ (or vice versa), then there exists
some $c\in (a,b)$ such that: 
\begin{equation*}
g(c)=y
\end{equation*}%
In our case, it requires checking limits at infinity: 
\begin{align*}
\lim_{u\rightarrow 0}J(u)& =\lim_{u\rightarrow 0}u-\alpha v\theta\tanh (-\gamma
u)=0  \\
\lim_{u\rightarrow 1}J(u)& =\lim_{u\rightarrow 1}u-\alpha v\theta\tanh (-\gamma
u)=1-\alpha v\theta\tanh (-\gamma )
\end{align*}%
Thus, by the IVT, for every $y\in \lbrack 0$,$1-\alpha v\theta\tanh(-\gamma
)]$, there exists $u\in \lbrack 0,1]$ such that $J(u)=y$. Therefore, $J(u)$
is surjective onto the interval $[0$,$1-\alpha v\theta\tanh(-\gamma )]$.

\subsection{Proof of Proposition 4}
To prove Proposition 4 using the existence part of the Andronov-Hopf bifurcation theorem, in this demonstration, we use $\beta_0$ as the bifurcation parameter. We must first (HB1) show that the characteristic equation possesses a pair of complex conjugate eigenvalues $\sigma\!\left(\beta_0\right) \pm i\,\omega\!\left(\beta_0\right)$ that become purely imaginary at the critical value $\beta_{0_{HB}}$ of the parameter, that is, $\sigma\!\left(\beta_{0_{HB}}\right)=0$, and that no other eigenvalues with zero real part exist at $\beta_{0_{HB}}$. Then, (HB2) we must check that the derivative of the real part of the complex eigenvalues with respect to the bifurcation parameter is nonzero at the
critical value.

(HB1) Given that the conditions $a_1 > 0, a_2 > 0$ and $a_3>0$ are all satisfied, in order that the characteristic equation has one negative real root and a pair of complex roots with zero real part, we must have:
\begin{equation*}
    a_1a_2-a_3=0
\end{equation*}
Making use of Eqs. (\ref{critical_condition}) and (\ref{e_equilibrium}), the condition above is satisfied for 
\begin{align*}
\bar{e}f'&=\beta_0+\alpha \{1-\theta [1+\tanh(-\gamma J^{-1}(1-\alpha v (1-\theta)))]\}=1
\end{align*}
(HB2) By using the so-called sensitivity analysis, it is then possible to show that the second requirement of the Andronov-Hopf bifurcation theorem is also met. Substituting the elements of the Jacobian matrix into the respective coefficients of the characteristic equation, as reported in the demonstration of Proposition 2, and making use of Eq. (\ref{e_equilibrium}), it follows:
\begin{eqnarray*}
    a_{1}&=&1\\
    a_{2}&=&\alpha \bar{u}\gamma\theta\left[ 1-\tanh ^{2}(-\gamma \bar{u})\right] +\frac{\bar{u}}{v}\{\beta_0+\alpha \{1-\theta [1+\tanh(-\gamma J^{-1}(1-\alpha v (1-\theta)))]\}\} \\
    a_{3}&=& \left\{\alpha \bar{u}\gamma\theta\left[ 1-\tanh ^{2}(-\gamma \bar{u})\right] +\frac{\bar{u}}{v}\right\}\{\beta_0+\alpha \{1-\theta [1+\tanh(-\gamma J^{-1}(1-\alpha v (1-\theta)))]\}\}
\end{eqnarray*}
so that
\begin{eqnarray*}
    \frac{\partial a_1}{\partial \beta_0}&=&0\\
    \frac{\partial a_2}{\partial \beta_0}&=&\frac{\bar{u}}{v}>0\\
    \frac{\partial a_3}{\partial \beta_0}&=& \alpha\bar{u}\gamma\theta\left[ 1-\tanh ^{2}(-\gamma \bar{u})\right]
    +\frac{\bar{u}}{v}>0
\end{eqnarray*}

When (HB1) holds, one root of the characteristic equation is real negative ($\lambda_1$), whereas the other two are a pair of complex roots with zero real part ($\lambda_{2,3}=\sigma\pm i\omega$, with $\sigma=0$). We thus have:
\begin{eqnarray*}
    a_1&=&-(\lambda_1+\lambda_2+\lambda_3)\\
    &=&-(\lambda_1+2\sigma)
\end{eqnarray*}
\begin{eqnarray*}
    a_2&=&\lambda_1\lambda_2+\lambda_1\lambda_3+\lambda_2\lambda_3\\
    &=& 2\lambda_1\sigma+\sigma^2+\omega^2
\end{eqnarray*}
\begin{eqnarray*}
    a_3&=&-\lambda_1\lambda_2\lambda_3\\
    &=& -\lambda_1(\sigma^2+\omega^2)
\end{eqnarray*}
such that:
\begin{eqnarray*}
    \frac{\partial a_1}{\partial \beta_0}&=&-\frac{\partial \lambda_1}{\partial \beta_0}-2\frac{\partial \sigma}{\partial \beta_0}=0\\
    \frac{\partial a_2}{\partial \beta_0}&=&2\sigma\frac{\partial \lambda_1}{\partial \beta_0}+2(\lambda_1+\sigma)\frac{\partial \sigma}{\partial \beta_0}+2\omega\frac{\partial \omega}{\partial \beta_0}=\frac{\bar{u}}{v}\\
    \frac{\partial a_3}{\partial \beta_0}&=& -(\sigma^2+\omega^2)\frac{\partial \lambda_1}{\partial \beta_0}-2\lambda_1\sigma\frac{\partial \sigma}{\partial \beta_0}-2\lambda_1\omega\frac{\partial \omega}{\partial \beta_0}\\
    &=& \alpha\bar{u}\gamma\theta\left[ 1-\tanh ^{2}(-\gamma \bar{u})\right]
    +\frac{\bar{u}}{v}
\end{eqnarray*}

For $\sigma=0$, the system to be solved becomes:
\begin{eqnarray*}
    -\frac{\partial \lambda_1}{\partial \beta_0}-2\frac{\partial \sigma}{\partial \beta_0}&=&0\\
    2\lambda_1\frac{\partial \sigma}{\partial \beta_0}+2\omega\frac{\partial \omega}{\partial \beta_0}&=&\frac{\bar{u}}{v}\\
    -\omega^2\frac{\partial \lambda_1}{\partial \beta_0}-2\lambda_1\omega\frac{\partial \omega}{\partial \beta_0}&=& \alpha\bar{u}\gamma\theta\left[ 1-\tanh ^{2}(-\gamma \bar{u})\right]
    +\frac{\bar{u}}{v}
\end{eqnarray*}
or
\[
\begin{bmatrix}
-1 & -2 & 0 \\
0 & 2\lambda_{1} & 2\omega \\
-\omega^{2} & 0 & -2\lambda_{1}\omega
\end{bmatrix}
\begin{bmatrix}
\dfrac{\partial \lambda_{1}}{\partial \beta_0} \\[6pt]
\dfrac{\partial \sigma}{\partial \beta_0} \\[6pt]
\dfrac{\partial \omega}{\partial \beta_0}
\end{bmatrix}
=
\begin{bmatrix}
0 \\[6pt]
\frac{\bar{u}}{v} \\[6pt]
\alpha\bar{u}\gamma\theta\left[ 1-\tanh ^{2}(-\gamma \bar{u})\right]+\frac{\bar{u}}{v}
\end{bmatrix}
\]

Thus:
\begin{eqnarray*}
\frac{\partial \sigma}{\partial \beta_0}
\Bigg|_{\beta_0 = \beta_{0_\mathrm{HB}}}
&=&
\frac{
\begin{vmatrix}
-1 & 0 & 0 \\
0 & \frac{\bar{u}}{v} & 2\omega \\
-\omega^{2} & \alpha\bar{u}\gamma\theta\left[ 1-\tanh ^{2}(-\gamma \bar{u})\right]+\frac{\bar{u}}{v}
 & -2\lambda_{1}\omega
\end{vmatrix}
}{
\begin{vmatrix}
-1 & -2 & 0 \\
0 & 2\lambda_{1} & 2\omega \\
-\omega^{2} & 0 & -2\lambda_{1}\omega
\end{vmatrix}
} \\
&=& \frac{(-1)(-2\lambda_1\omega\frac{\bar{u}}{v}-2\omega\alpha\bar{u}\gamma\theta[1-\tanh^2(-\gamma\bar{u})]-2\omega\frac{\bar{u}}{v})}{4\omega(\lambda_1^2+\omega^2)}\\
&=& \frac{2\omega(\lambda_1\frac{\bar{u}}{v}+\frac{\bar{u}}{v}+\alpha\bar{u}\gamma\theta[1-\tanh^2(-\gamma \bar{u})])}{4\omega(\lambda_1^2+\omega^2)}\\
&=& \frac{2\omega((1+\lambda_1)\frac{\bar{u}}{v}+\alpha\bar{u}\gamma\theta[1-\tanh^2(-\gamma \bar{u})])}{4\omega(\lambda_1^2+\omega^2)} >0
\end{eqnarray*}


\newpage

\bibliography{references}

\end{document}